\documentclass[showpacs,aps]{revtex4}
\usepackage{graphicx}

\begin{document}  

\title{ Sphalerons, knots, and dynamical compactification in
Yang-Mills-Chern-Simons theories}
\author{John M. Cornwall\footnote{Email cornwall@physics.ucla.edu} and
Noah Graham\footnote{Email graham@physics.ucla.edu}}
\affiliation{Department of Physics and Astronomy, University of California,
Los Angeles, CA 90095}

\begin{abstract}
\pacs{  11.15.Tk, 11.15.Kc  \hfill UCLA/02/TEP/6}
Euclidean $d=3$ $SU(2)$ Yang-Mills-Chern-Simons (YMCS) theory, including
Georgi-Glashow (GGCS) theory, may have solitons
in the presence of appropriate mass terms.
For integral CS level $k$ and for solitons carrying integral CS number
$N_{CS}$, YMCS is gauge-invariant and consistent, and the CS integral
describes the compact Hopf map $S^3\rightarrow S^2$.  However,
individual solitons such as
sphalerons and linked center vortices with $N_{CS}=1/2$ and writhing center
vortices with arbitrary real $N_{CS}$ are non-compact; a condensate of them threatens  compactness of the theory.  We study
various forms of the non-compact theory in the dilute-gas approximation, 
including   odd-integral or non-integral values for the CS level $k$, treating the
parameters of non-compact large gauge transformations as collective
coordinates.  Among our conclusions:  1)  YMCS theory dynamically
compactifies; a putative non-compact YMCS theory has infinitely higher
vacuum energy $\int d^3x\epsilon_{vac}$ than compact YMCS.  2)  For
sphalerons with $N_{CS}=1/2$, compactification arises through
a domain-wall sphaleron, a pure-gauge configuration lying on a closed
surface carrying the right amount of $N_{CS}$ to compactify.  3)  We
can interpret the domain-wall sphaleron in terms of fictitious closed
Abelian field lines, associated with an Abelian potential and magnetic
field derived from the non-Abelian CS term.  In this language,
sphalerons are under- and over-crossings of knots in the field lines;
a domain-wall sphaleron acts as a superconducting surface which
confines these knots to a compact domain.  4)  Analogous results hold
for the linking and writhing of center vortices and nexuses.  5)  If
we induce a CS term with an odd number of fermion doublets,
domain-wall sphalerons are related to non-normalizable fermion zero
modes of solitons.  6)  GGCS with monopoles is explicitly compactified
with center-vortex-like strings.
\end{abstract}
 
\maketitle

\section{Introduction}

Compactification of  Euclidean space $R^d$, such as $R^d\rightarrow
S^d$, famously leads to integral quantization of certain topological
charges, such as the usual four-dimensional topological charge
commonly associated with instantons.  Yet practically since instantons
were invented, there have been indications of fractional topological
charge \cite{ffs,th81,co95,cy,co98,vb,ehn,co99,er00,co00,co02}, whose
existence could interfere with compactification. The basic issue we
address in this paper is whether compactification for Euclidean $d=3$
$SU(2)$ Yang-Mills-Chern-Simons (YMCS) theory is a mathematical
hypothesis, which could be abandoned, or whether there are dynamical
reasons for expecting it.  If the theory has either a Chern-Simons (CS) level
$k$ less than a critical value $k_c\simeq (2\pm 0.7)N$ for gauge group $SU(N)$ \cite{co96}, or a
fundamental Higgs field, there can exist solitons with CS number of
$1/2$, such as sphalerons and distinct linked center
vortices, or solitons with arbitrary real CS number, such as writhing
center vortices.  A condensate of such solitons, taken naively,
violates compactness and, if it has an interpretation at all, requires
integrating over all non-compact gauge transformations as collective
coordinates.  We find that candidate vacua in the dilute-gas
approximation have the lowest energy when the total $N_{CS}$ is
integral and $R^3$ is compactified to $S^3$. We find an interpretation
for this dynamical compactification in terms of a domain-wall
``sphaleron''  supplying  enough fractional CS number to compensate for
the total CS number of the bulk solitons.

In other circumstances, such as for fractional $d=4$ topological
charge, dynamical compactification apparently occurs, sometimes
through \cite{ct} the formation of a string that joins enough
fractional objects so that their total charge is integral. Or
compactness of center-vortex sheets may ensure \cite{er00,co00,co02}
topological confinement of $d=4$ topological charge.  Pisarski
\cite{pis} claims that TP monopoles in Georgi-Glashow theory with a CS
term added (GGCS theory) are joined by strings.  Any attempt to
separate out a half-integral set of topological charges joined by
strings requires the introduction of enough energy to stretch and
ultimately break the string. This in turn supplies new fractional
topological charges that enforce compactification.   Affleck {\it et
al.} \cite{ahps} argue that in CCGS the long-range monopole fields
already decompactify the space and allow arbitrary CS number; summing
over these arbitrary values leads to suppression of TP monopoles.  For
the condensed-matter analog, see Ref. \cite{fs}. (Ref. \cite{hts}
challenges the conclusion of \cite{ahps}, on different grounds, which
are somewhat related to ours.)  More recent developments require
modifications of the views of \cite{pis,ahps}.  Several authors
\cite{co98,ag,fgo} have pointed out that the TP monopole is actually
like a nexus, with its magnetic flux confined into tubes that are,
for all practical purposes, the tubes of center vortices.  These
tubes, which join a monopole to an anti-monopole, are the natural
candidates for the strings Pisarski \cite{pis} claims to exist,
although his claim is not based on finding any stringlike object in CCGS.  We find no evidence for strings joining sphalerons, and none is expected since if every $d=3$ cross-section of $d=4$ space had an even number of sphalerons, the $d=4$ space would have to have topological charge which is an even integer.  But there is no such restriction in $d=4$.

In this paper we will be concerned with a  YMCS theory with extra mass
terms, which can either be due to quantum effects or added
explicitly.  For YMCS theory with no explicit mass terms it has been
argued \cite{co96} that if the CS level $k$  is less than a critical
value $k_c$, the CS-induced gauge-boson mass is not large enough to
cure the infrared instability of the underlying YM theory.  In this
case a dynamical mass (equal for all  gauge bosons) is generated just
as for the YM theory with no CS term. In YM theory it is known
that quantum sphalerons of CS number $1/2$ exist \cite{co76}, as well
as center vortices with various CS numbers.  Ref. \cite{co96} estimates
$k_c\simeq (2\pm 0.7)N$ for gauge group $SU(N)$.  But for $k>k_c$, YMCS
with no matter fields is essentially perturbative, and in the same
universality class as Witten's  topological gauge theory \cite{wit89}
which has only the CS term in the action; there are no solitons of finite
action \cite{dv}.   There are also sphalerons \cite{dhn} and center
vortices for YM theory with an elementary Higgs field in the
fundamental representation; this theory is quite close in behavior to
YM theory with no matter fields and $k<k_c$.
 
A fundamental assumption of the present paper is that if these
isolated  sphalerons or center vortices exist, a condensate of them is
allowed, and that we can learn something about this condensate through
conventional dilute-gas arguments.  This assumption could fail if the
sphalerons are somehow so strongly coupled that the dilute-gas approximation is qualitatively wrong, but it is not our purpose
to investigate this possibility.  Certainly, lattice evidence for a
center vortex condensate in QCD suggests that there is some sense to
the dilute-gas approximation.

The sphaleron \cite{dhn,co76,mk} is the prototypical example of an
isolated non-compact soliton in $d=3$ gauge theory; it nominally has
Chern-Simons number ($N_{CS}$) of $1/2$, instead of the integral value
demanded by compactification.  No apparent problems arise until one
adds a CS term  \cite{djt,sch}.  For odd CS level $k$,
an odd number of sphalerons is detectable in the partition function
and elsewhere as a non-compact object, which seems to break gauge
invariance and in any event leads to peculiar signs for physical
objects.  In this hybrid theory with integral and odd CS level $k$, we
do not admit arbitrary non-compact gauge transformations, but we do
admit a condensate of sphalerons, each with $N_{CS}=1/2$.  In sectors
with an odd number $J$ of sphalerons, the total CS number $J/2$ is
half-integral and non-compact.  We argue that it is energetically
favorable to form a domain-wall ``sphaleron,'' which may live on the
surface at infinity and which also carries half-integral CS number.
The total CS number of the explicit sphalerons and the domain wall is
now integral, its energy is lowered, and the theory is dynamically
compactified.

Next we explore the consequences of  admitting not only non-compact
solitons but also non-compact gauge transformations.  Then the CS
number of any isolated $d=3$ gauge-theory soliton is essentially
arbitrary, since it can be changed by a large non-compact gauge
transformation of the form
\begin{equation}
\label{largegt}
U=\exp [i\alpha (r)\vec{\tau}\cdot\hat{r}/2];\;\;\alpha (\infty )\neq 2\pi N.
\end{equation}
This gauge transformation does not change the action (aside from the CS term itself) or equations of
motion of the soliton, and therefore corresponds to a collective
coordinate.  As we will see, integrating over $\alpha (\infty )$ as a
collective coordinate does not automatically compactify the space; it simply
gives it a higher vacuum energy density (and infinitely higher energy
for infinite volume) than it would have if the solitons allowed
compactification.  Again we expect the energy of the vacuum to be
lowered by formation of a domain-wall sphaleron.

\begin{figure}
\includegraphics{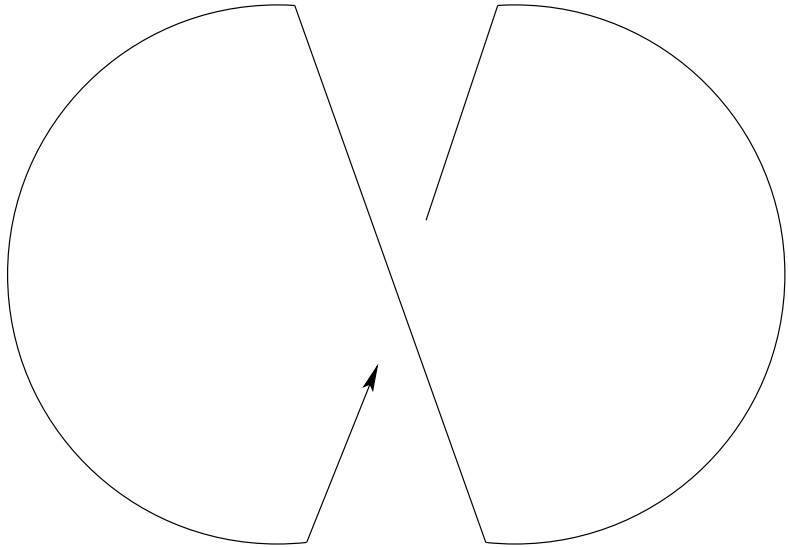}
\caption{\label{fig1}A center vortex with writhe}
\end{figure}

The sphaleron is not the only object in $SU(2)$ $d=3$ YM or YMCS
theory with half-integral CS number.  In the center vortex-nexus view
of gauge theories \cite{th78,co79,mp,no,cpvz,bvz} $d=4$ topological
charge is carried (in part) by the linkage of center vortices and
nexuses \cite{er00,co00,co02}; the $d=3$ projection of such linkages
is that center vortices, which are closed fat strings of magnetic
flux, carry CS number through mutual linkage as well as the
self-linkage of twisting and writhing
\cite{co95,co96,co02,engel,bef,rein} of these strings.  In their
simplest linked configuration, which consists of two untwisted but
linked loops whose distance of closest approach is large compared to
the flux-tube thickness, they carry a mutual link number $Lk$,
which is an integer, and a CS number of $Lk/2$, just like a sphaleron.
But there are isolated configurations that carry essentially any link or
CS number \cite{co95,cy}.  These are individual center vortices with
writhe.  An example is shown in Fig. \ref{fig1}.  For
mathematically-idealized (Dirac-string) vortices, one can apply
Calugareanu's theorem, which says that the ribbon-framed self-linking
number $FLk$ is an integer and a topological invariant, although not
uniquely defined, and that $FLk=Tw+Wr$ where the writhe $Wr$ is the
standard Gauss self-linking integral and $Tw$ is the twist or torsion
integral.  Neither $Tw$ nor $Wr$ is a topological invariant, and
neither is restricted to take on integer values.  Since the CS
number is proportional to the writhe, even
without making general non-compact gauge transformations, one has the
phenomenon of essentially arbitrary CS number for an isolated soliton.
The effect of these center vortices is much the same as if we admit
non-compact gauge transformations.  One difference is that for
spatially-compact center vortices the CS density is localized, while
for non-compact gauge transformations the associated CS density always
lies on the surface at infinity.

For physical center vortices, where Dirac strings turn into fat tubes
of magnetic flux, the same picture holds, although for different reasons.
For such fat tubes, the standard integrals for $Tw$ and $Wr$ are
modified, and there is no sharp distinction between twist and writhe
\cite{co95}.  The modified integrals depend on the details of the
mechanism that fattens the flux tubes (which may be thought of as the
generation of a dynamical mass by quantum effects arising from
infrared instability of the YM theory).  There is no reason for $Tw$
or $Wr$ to be integers, or even fractions such as $1/2$.  In physics
language,  $Tw$ and $Wr$ are dependent on collective
coordinates of the vortex.  So the situation should be somewhat
similar to that for sphalerons.

If CS numbers can take on arbitrary values, how can $d=4$ topological
charge, which is the difference of CS numbers, be restricted to
integral values?  The answer, of course, is $d=4$ compactness, which is
not related to $d=3$ compactness; a non-compact $d=4$ space may have only
compact $d=3$ cross-sections.  Compactness in $d=4$ constrains the CS
numbers whose difference is the topological charge, so
that not every pair of $d=3$ condensates of CS number can occur in a
compact $d=4$ space.  We give some simple examples, analogous to the
pairing of crossings (sphalerons) in a compact $d=3$ space, to
show how these constraints arise.  These examples interpret the net change
in CS number as arising from a dynamic reconnection process, in which
center vortices change their link number; only certain kinds of
dynamic reconnection are allowed by compactness. The point of
reconnection, when two otherwise distinct vortices have a common
point, is simply the point of intersection of center vortices in the
previously-studied picture of $d=4$ topological charge as an
intersection number of closed vortex surfaces (plus linkages of these
surfaces with nexus world lines) \cite{er00,co00,co02}; such
intersection points come in pairs for compact vortex surfaces.
    
We summarize our results as follows:
\begin{enumerate}
\item
Even if compactification is not assumed {\it a priori}, and
solitons possess collective coordinates amounting to arbitrary
$N_{CS}$ for every soliton, the lowest-energy candidate vacuum state
of a YMCS theory is one which is compact.  The energetic favorability
of compactification we describe by  {\em dynamical compactification}.
\item 
There is no evidence for strings that locally bind sphalerons
into paired objects; instead, there is evidence for a domain-wall
``sphaleron'' carrying half-integral CS number if the bulk
sphalerons also carry half-integral CS number.
\item
Sphalerons can be mapped onto over- and under-crossings of knots 
occuring in closed fictitious Abelian field lines  associated
with the non-Abelian CS term.  There is always an even number of
crossings for compact knots, and so an odd number of crossings of
closed knot components must be compensated by an odd number of crossings
elsewhere.  The domain-wall sphaleron acts as a superconducting wall
that confines the closed fictitious field lines to a compact domain,
where they must close and have an integral CS number.
\item
Fat self-linked center vortices can also carry arbitrary $N_{CS}$
associated with collective coordinates; we expect phenomena similar to
those found for sphalerons.
\item
For a compact $d=4$ condensate of center vortices and nexuses, living in
the product of a $d=3$ space and a (Euclidean) ``time'' variable,
possible time-dependent  reconnection of vortices that would
change link numbers is constrained by compactness to yield integral
topological charge.
\item
If fermions are added, we show that half-integral fermion number and
half-integral CS number go together, and identify an extra fermion
number of $1/2$ with a fermion zero mode at infinity, which is a zero
mode of the domain-wall sphaleron at infinity.  This result
generalizes to the case of arbitrary CS number as well.
\item
In the GGCS model, strings exist that bind TP monopoles to TP
anti-monopoles and restore compactification; these strings are
essentially those of center vortices, while the TP monopoles are like
nexuses.
\end{enumerate} 

\section{Solitons in YMCS theory}
\label{solitons} 

In this section we establish notation, review the properties of
solitons with the usual spherical {\it ansatz} in massive YMCS
theory, and remark that every configuration in the functional
integral for YMCS theory has a conjugate related by a Euclidean
CPT-like transformation.  
 
\subsection{Spherically-symmetric solitons of YMCS theory}

The action of YMCS is complex, so its classical solitons can be
complex too.  Then the CS action $I_{CS}$ defined in
Eq. (\ref{csaction}) below may be complex, and its real part is not
interpretable as a CS number.  We will always define $Im\;
I_{CS}=2\pi k N_{CS}$, where the integer $k$ is the CS level.
In general a large gauge transformation only changes the imaginary
part of $I_{CS}$, and so this identification makes sense.   But one
may also ask whether it makes sense at all to discuss complex solitons
as extrema of the action; certainly, as Pisarski \cite{pis} points
out, this is quite wrong in some circumstances.  The other possibility
is to use only solitons of the real part of the action, and simply
evaluate the CS term at these solitons.  Of course, this works fine in
$d=4$, where the theta term adds nothing to the equations of motion.
Ref. \cite{co96} argues that in YMCS there is a complex (but
self-conjugate) spherically-symmetric soliton much like a sphaleron;
the particular case studied there had purely real CS action and hence
no CS number.   We show here that the would-be sphaleron of
Ref. \cite{co96} can easily be promoted to a sphaleron with  $N_{CS}$=
$1/2$.

One knows \cite{co82, chk} that with no CS term, $d=3$ YM theory with
no matter terms is infrared-unstable and non-perturbative, requiring
the dynamical generation of a gluon mass $M$ of order $Ng^2$ for gauge
group $SU(N)$, where $g$ is the gauge coupling.  If this theory is extended
to YMCS theory, it appears (at least from one-loop calculations \cite{co96}) that
the Chern-Simons gauge-boson mass $m=kg^2/4\pi$ is too
small to cure the infrared instability, and so generation of dynamical
mass is still required.  The estimates of the critical level $k_c$ are
based on one-loop calculations of the gauge-invariant pinch-technique
(PT) gauge-boson propagator \cite{co82,chk,cp,papa} and may not be
very accurate, but unpublished estimates of two-loop corrections by
one of us (JMC) suggest that the existence of a finite $k_c$ is
well-established.  The one-loop calculations give $k_c \simeq 2-5$ for $SU(2)$.  The generation of a dynamical mass generally leads to
confinement, via the creation of a condensate of center vortices
\cite{th78,co79,mp,no} and nexuses
\cite{co98,co99,er00,co00,cpvz,bvz}.  The long-range effects essential
for confinement come from pure-gauge parts that disorder the Wilson
loop ({\it i. e.}, give it an area law) by fluctuations in the Gauss
linking number of vortices and the Wilson loop. 

Define the usual anti-Hermitean $SU(2)$ gauge-potential matrix with
the gauge coupling $g$ incorporated by:
\begin{equation}
\label{notation}
A_j(\vec{x})=(\frac{g}{2i})\tau_aA_j^a(\vec{x})
\end{equation}
where the component form $A_j^a(\vec{x})$ is the canonical gauge
potential.   The Euclidean YM action is:
\begin{equation}
\label{action}
I_{YM}=\int d^3x  \frac{-1}{2g^2}TrG_{ij}^2 .
\end{equation}
To this can be added the Chern-Simons action:
\begin{equation}
\label{csaction}
I_{CS}=(2\pi i k)Q_{CS};\;\;Q_{CS}=\frac{-1}{8\pi^2}\int
d^3x\epsilon_{ijk}Tr[A_i\partial_jA_k+\frac{2}{3}A_iA_jA_k].
\end{equation}
The sum $I_{YM}+I_{CS}$ is the YMCS action $I_{YMCS}$.
Throughout this paper we will define the CS number $N_{CS}$ as the
{\em real} part of the integral in Eq. \ref{csaction}:
\begin{equation}
\label{ncs}
N_{CS}\equiv Re\;Q_{CS}=\frac{Im\;I_{CS}}{2\pi k}.
\end{equation}
It is only from this real part that phase or gauge-invariance problems
can arise.  Gauge invariance under large (compact) gauge
transformations requires that the Chern-Simons level $k$ is an
integer, so that   the integrand $\exp -I_{CS}$ of the partition
function is unchanged.  At the classical level, all gauge bosons
acquire a Chern-Simons mass $m\equiv kg^2/4\pi$.

As mentioned in the introduction, the CS mass may not be large enough
to cure the infrared instabilities of 
YMCS with no matter fields, and a dynamical mass is generated.  This
mass is the same for all gauge bosons.  The infrared-effective action
for this dynamical mass \cite{co82} is just a gauged non-linear sigma model:
\begin{equation}
\label{dynmass}
I_M=\frac{-m^2}{g^2}\int d^3x Tr[ U^{-1}D_iU]^2;\;\;
D_i=\partial_i+A_i;\;\;U=\exp (i\omega_a\tau_a/2).
\end{equation} 
When the unitary matrix $U$ and the gauge potential have the following
gauge-transformation laws, the action $I_M$ is gauge-invariant:
\begin{equation}
\label{gauge}
U\rightarrow VU;\;\;A_i\rightarrow VA_iV^{-1}+V\partial_iV^{-1}.
\end{equation}
The effective action $I_{eff}\equiv I_{YMCS}+I_M$ is valid in the
infrared regime, but at large momentum $p$ or short distance $x$ the
dynamical mass $m^2$ necessarily vanishes at a rate $\sim p^{-2}$ or
$x^2$ (modulo logarithms). This dynamical-mass effective action is the
same, for our purposes, as if one added a fundamental Higgs field, as
in the Weinberg electroweak action.

Because the action is complex, in general we must deal with complex
values for the gauge potentials and matter fields.  However, the
matrix $U$ must always be an $SU(2)$ matrix; that is, in the component
form $U=\exp (i\tau_a\omega_a/2)$ the fields $\omega_a$ are always
real.

\subsection{Complex field configurations}

With a complex action, there is no reason to restrict the path
integral to real fields.  There is an elementary theorem, essentially
a Euclidean CPT theorem, applicable to complex YMCS gauge fields and
any scalar fields, such as the fields $\psi(x)$ of the GG model
discussed later.   Given any configuration of gauge and scalar fields
for which the actions evaluated on
this configuration have the values $I_{YM},I_{CS},I_M,I_{GG}$, we
define a conjugate configuration by:   
\begin{equation}
\label{cpt}
CPT:\;\; A_i(x),G_{ij}(x),U(x),\psi (x)\rightarrow 
A^{\dagger}(-x),-G_{ij}^{\dagger}(-x),U(-x),\pm \psi (-x)^{\dagger} 
\end{equation}
or in component language:  
\begin{equation}
A_i^a(x)\rightarrow -A_i^a(-x)^*,\omega_a(x)\rightarrow
\omega_a(-x),\phi_a(x)\rightarrow \mp \phi_a(-x)^*.
\end{equation}
The possible sign change of the scalar field $\psi$ reflects any intrinsic parity.
Then the CPT-transformed configuration has actions
$I^*$, $I_{CS}^*$, $I_M^*$, and $I_{GG}^*$.  Note that
$N_{CS}$ changes sign under conjugation.

Below we will look for solitons of the YMCS action plus matter terms.
Generally these solitons, like the action itself, will be complex.
They can be divided into two types: 1) those configurations
  that transform
into themselves under CPT, which we call self-conjugate,  and 2) those
that transform to another configuration.  Self-conjugate
configurations  have {\em real} action, including the CS term.  It is
easy to see that if any configuration of type 2)  satisfies the
complex equations of motion then so does its CPT conjugate, and both are admissible solitons if
either is. Examples of type 1) solitons are given in \cite{co96}, for
the YMCS action with dynamical mass generation.  These solitons cannot
be said to possess topological properties as expressed through the CS
term, since the CS number $N_{CS}$ vanishes. However, from this
 self-conjugate soliton it is easy to generate solitons that are not self-conjugate with any desired CS number.

We review the sphaleron-like complex soliton \cite{co96} of the action
$I_{YMCS}+I_M$ (see equations (\ref{action}, \ref{dynmass})).  Using
the notation of \cite{co96}, a spherical soliton is described by four
functions of $r$:
\begin{equation}
\label{sphera}
2iA_i=\epsilon_{iak}\tau_a\hat{x}_k(\frac{\phi_1(r)-1}{r}) -
(\tau_i-\hat{x}_i\hat{x} \cdot
\vec{\tau})\frac{\phi_2(r)}{r}+\hat{x}_i\hat{x}\cdot \vec{\tau}H_1(r),
\end{equation}
\begin{equation}
\label{spheru}
U=\exp (i\beta (r)\frac{\vec{\tau}\cdot \hat{x}}{2}).
\end{equation}
The equations of motion, found by varying both $A_i$ and $U$, are:
\begin{equation}
\label{eom1}
0=(\phi_1^{\prime}-H_1\phi_2)^{\prime}+\frac{1}{r^2}\phi_1(1-\phi_1^2-\phi_2^2)+(im-H_1)(\phi_2^{\prime}
+H_1\phi_1)-M^2(\phi_1-\cos \beta );
\end{equation}
\begin{equation}
\label{eom2}
0=(\phi_2^{\prime}+H_1\phi_1)^{\prime}+\frac{1}{r^2}\phi_2(1-\phi_i^2-\phi_2^2)-(im-H_1)(\phi_1^{\prime}
-H_1\phi_2)-M^2(\phi_2+\sin \beta );
\end{equation}
\begin{equation}
\label{eom3}
0=\phi_1\phi_2^{\prime}-\phi_2\phi_1^{\prime}+H_1(\phi_1^2+\phi_2^2) +
(im(1-\phi_1^2-\phi_2^2) +\frac{1}{2}M^2r^2(H_1-\beta^{\prime});
\end{equation}
\begin{equation}
\label{eom4}
0=\frac{1}{r^2}[r^2(\beta^{\prime}-H_1)]^{\prime}-\frac{2}{r^2}(\phi_1\sin
\beta + \phi_2 \cos \beta )
\end{equation}
where
\begin{equation}
\label{csmass}
m=\frac{kg^2}{4\pi}
\end{equation}
is the Chern-Simons mass at level $k$ and the prime signifies
differentiation with respect to $r$.  These equations   reduce to
those of \cite{co96} at $\beta = \pi$.  As in \cite{co96},
Eq. (\ref{eom4}), which is the variational equation for $U$, is not
independent of the other three equations. It can be derived from them
by simple manipulations because there is still an Abelian gauge degree
of freedom:
\begin{eqnarray}
\label{abelg}
\phi_1(r) \rightarrow \phi_1(r) \cos \alpha(r)+\phi_2(r) \sin
\alpha(r) &\quad& \phi_2(r) \rightarrow \phi_2(r) \cos \alpha(r) -
\phi_1(r) \sin \alpha(r)\cr
\beta (r) \rightarrow \beta (r) + \alpha (r) &\quad&
H_1(r)\rightarrow H_1(r) +\alpha^{\prime}(r).
\end{eqnarray}
The boundary conditions are:
\begin{eqnarray}
\label{bc}
r=0: & \phi_1(0)=1;\;\;\phi_2(0)=H_1(0)=\beta (0)=0; \\ \nonumber
r=\infty : & \;\;\phi_1(\infty)=\cos \beta (\infty
);\;\;\phi_2(\infty) = -\sin \beta (\infty ).
\end{eqnarray}

First consider the case $\beta = \pi$.  Then \cite{co96} there is a
solution where $\phi_1$ is real and $\phi_2$ and $H_1$ are pure
imaginary.  This corresponds to a self-conjugate soliton, so the CS
action is purely real (that is, the CS integral $Q_{CS}$ in Eq.
(\ref{csaction}) is pure imaginary). This is easily checked from the
explicit form
\begin{equation}
\label{spherncs}
Q_{CS}=\frac{1}{8\pi^2}\int
\frac{d^3x}{r^2}[\phi_1\phi_2^{\prime}-\phi_2\phi_1^{\prime} -
\phi_2^{\prime}-H_1(1-\phi_1^2-\phi_2^2)].
\end{equation}

If any solution of the equations of motion is gauge-transformed as in
Eq. (\ref{abelg}), it remains a solution to these equations and all
contributions to the action are unchanged except, of course, for the
CS part of the action.  If we start with the self-conjugate soliton
above, and transform it with a function $\alpha (r)$ such that $\alpha
(0)=-\pi ,\;\alpha (\infty )=0$ one sees that the soliton is no longer
self-conjugate, and in general all three functions $\phi_{1,2},\;H_1$
are complex.  This choice of boundary conditions for $\alpha$
removes an integrable singularity in the original self-conjugate
sphaleron, but does not change the YM and mass parts of the action.
The change in the CS integral, because it does not affect the
equations of motion,  is necessarily a surface term:
\begin{equation}
\label{qcs}
\delta Q_{CS}=\frac{1}{2\pi}\left[\alpha (r)-\sin \alpha (r)
\right|_0^{\infty} = \frac{1}{2}.
\label{Qcseq}
\end{equation}
The new sphaleron has $N_{CS}=1/2$, as appropriate for a sphaleron.

The immediate objection is that one could as well choose any value for
$\alpha (\infty )$, and change the sphaleron's CS number to any
desired value.  Integration over this collective coordinate might
cause sphalerons to be confined in pairs (as argued in \cite{ahps} for
TP monopoles in the GGCS model).  However, it does not quite happen
that way for sphalerons.  We next show that integrating over $\alpha
(\infty )$ for all sphalerons does increase the free energy, but does
not lead immediately to confinement of sphalerons in pairs.  In such a
case, compactification becomes  the preferred state dynamically.
 
\section{Dynamical compactification}
\label{puzzle} 

As discussed in the introduction, sphalerons (and center vortices)
present a challenge to the usual view of compact YMCS, since these
solitons in isolation violate compactness and lead to problems with
gauge invariance.  In this section we consider several cases,
beginning with the internally-inconsistent but instructive case in
which $k$ is integral and only compact gauge transformations area
allowed but there is a condensate of (non-compact) sphalerons.   For odd $k$ the energy density of the vacuum is changed
in sign from the case of even $k$, which raises the vacuum energy by
an infinite amount.    In the next case, $k$ is still an integer but
we allow large gauge transformations of the form $\exp [i\vec{\tau}\cdot
\hat{r}\alpha (r)/2]$ with arbitrary $\alpha (\infty )$.  Since the
action of a sphaleron depends on $\alpha (r)$ only through the CS
phase factor, this variable can be treated as a collective coordinate
and integrated over.  We will see that this integral again raises the
free energy, suggesting that the compactified theory is preferred on
energetic grounds.  Finally, we consider the case of general $k$,
including spatially-variable $k$, and non-compact gauge
transformations and find, analogous to L\"uscher's work \cite{lus} in
$d=4$, that if $k$ takes on a non-integral value in a bounded domain
and an integral value outside it, this domain or ``bag'' has a positive
energy, scaling with the bag volume, above the integral-$k$ vacuum.
 
This last case gives us a clue to what actually causes the ostensibly
non-compact theory to compactify.  We find no evidence for strings
that would join pairs of sphalerons together, nor do the collective
coordinate integrations reduce the theory to the zero-sphaleron
sector.   Instead, we argue in Sec. \ref{knotsec} that among the
collective coordinates for large gauge transformations with any value
of $\alpha (\infty)$, there is the possibility of formation of a
domain-wall sphaleron that places half-integral CS number on a closed
surface surrounding an odd number of sphalerons, to add to the
half-integral CS number present from the sphalerons inside.   This
domain wall itself has no energy, and is a pure-gauge object; it can
be moved around, deformed, and so on, without changing the physics.  It
acts as a superconducting wall that causes the fictitious Abelian
field lines associated with non-Abelian CS number to be confined to
the interior of the domain wall, or in other words to be compact.

Our arguments are based on the assumption that a condensate of
sphalerons in YMCS theory can be treated in the dilute-gas
approximation, or equivalently that all solitons are
essentially independent.  When a CS term is present in the action, the
partition function $Z$ is the usual expansion as a sum over sectors of
different sphaleron number:
\begin{equation}
\label{z}
Z(k)=\sum_JZ_J;\;\;Z_J(k)= \sum_{c.c.}\frac{1}{J!}e^{-\sum I_c}+\dots
\end{equation}
where $Z_J(k)$ is the partition function in the sector with $J$
sphalerons; the subscript $c.c.$ indicates a sum over collective
coordinates of the sphalerons; $I_c$ is the action (including CS
action) of a sphaleron and the omitted terms indicate corrections to
the dilute-gas approximation.  To be more explicit, we separate the
sum over collective coordinates into kinematic coordinates, such as
spatial position, and gauge collective coordinates.  The former we
represent in the standard dilute-gas way and the latter we indicate as a
functional integral over large gauge transformations $U$: 
\begin{equation}
\label{zcc}
Z(k)=\int (dU) \sum\frac{1}{J!}(\frac{V}{V_c})^J 
\exp -\{J Re\;I_c+2\pi ik[JN_{CS}(A_c)+N_{CS}(U)]\}
\end{equation}
Here $Re\;I_c$ is the real part of the action, $N_{CS}(A_c)$ is the CS
number of each individual soliton of gauge potential $A_c$ (taken in
some convenient gauge), and $N_{CS}(U)$ is the CS number of the large
gauge transformation.  As in Sec. \ref{solitons}, we choose
$A_c$ so that $N_{CS}(A_c)=0$.

If we now restrict the large gauge transformations $U$ to be compact,
so that $N_{CS}(U)=K$, an integer, we recover the standard \cite{djt}
result that $Z(k)$ is non-zero only for $k$ an integer.

Now retain the assumption that only compact gauge transformations are
allowed and that $k$ is integral, but allow a condensate of
sphalerons.  Sphalerons correspond to a limitation of $\alpha
(a;\infty )$ to the two values $\pm \pi$.   The $\sin \alpha$ term vanishes
in Eq. (\ref{Qcseq}), and the collective-coordinate sum reduces to:
\begin{equation}
\label{zsphal}
Z=\sum_{J_+,J_-}\frac{1}{J_+!J_-!}(\frac{V}{V_c})^J \exp
-[JRe\;I_c]e^{ik\pi (J_+-J_-)}=\exp \{e^{i\pi k}2(\frac{V}{V_c})
\exp -[Re\;I_c]\}   ;\;\;(J=J_++J_-).
\end{equation}
If $k$ is odd, this expression for $Z$ has precisely the opposite sign
in the exponent to that of a normal dilute-gas condensate, which means
that the free energy, which for a normal dilute-gas condensate is
negative, has turned positive.  So the non-compactified theory
has in a higher free energy than the compactified
theory.  (Non-compactification also leads to a number of
other unphysical results in the dilute-gas approximation, which we
will not dwell on here.)
   
Now consider the case of non-compact gauge transformations.
Suppose that, as in Sec. \ref{solitons}, the sphalerons
are obtained by a gauge transformation of the
form given in Eq. (\ref{abelg}) acting on a self-conjugate soliton,
whose action is real and positive.  The $a^{th}$ soliton is at
position $\vec{r}-\vec{a}\equiv \vec{r}(a)$.  Denote by $\alpha
(a;\infty )$ the asymptotic value of the gauge variable for the
$a^{th}$ soliton.  Since the total CS number of all $J$ sphalerons
comes from a surface contribution, we can immediately write the phase
factor in the action by generalizing Eq. (\ref{qcs}):
\begin{equation}
\label{z2}
Z(k)=\sum_J\frac{1}{J!}(\frac{V}{V_c})^J\exp -[J Re\;I_c]
\exp ik[\alpha -\sin \alpha ]
\end{equation}
where
\begin{equation}
\label{alphasum}
\alpha = \sum_{a=1}^J\alpha (a;\infty ).
\end{equation}
 
We are treating the $\alpha (a;\infty )$ as collective coordinates, so
we integrate over them:
\begin{equation}
\label{ccint}
Z(k)=\sum_JZ_{RJ} \times \{\prod_a \int_0^{2\pi}
\frac{d\alpha(a;\infty )}{2\pi} \}\exp ik[\alpha -\sin \alpha ]
\end{equation} 
where $Z_{RJ}$ indicates the explicitly real terms in the summand of
Eq. (\ref{z2}).  This integral is reduced to a product by using the
familiar Bessel identity
\begin{equation}
\label{bessel}
e^{ iz\sin \theta} \equiv \sum_{-\infty}^{\infty}J_N(z)e^{iN\theta}
\end{equation}
with the result, for integral $k$, $[J_k(k)]^J$.  So the dilute-gas
partition function is:
\begin{equation}
\label{z3}
Z(k)=\sum_J\frac{1}{J!}(\frac{V}{V_c})^J\exp J[-Re\;I_c+\ln J_k(k)] =
\exp \{\frac{V}{V_c}e^{-Re\;I_c}J_k(k)\}.
\end{equation}
Since $1\geq J_k(k)>0$ for all levels $k$, we see that integrating
over the collective coordinates has increased the free energy
(the negative logarithm of $Z$). This suggests that by
properly compactifying the sphalerons, so that the gauge behavior at infinity
is under control, we will lower the free energy, yielding
something like the usual dilute-gas partition function (which is
Eq. (\ref{z3}) without the $J_k(k)$ factor). 

Once one allows non-compact gauge transformations one might as well
allow non-integral $k$.  The results are analogous to those found long
ago by L\"uscher \cite{lus} for $d=2$ $CP^N$ models and $d=4$ gauge
theory with instantons and a $\theta$ angle.  Of course, the
calculations for non-integral $k$ only make sense in the non-compact
case.   For non-integral $k$ the function $J_k(k)$ of Eq. (\ref{z3})
must be replaced by
\begin{equation}
\label{nonint}
F(k)= \sum_{-\infty}^{\infty}J_N(k)\frac{\sin [\pi (k-N)]}{\pi (k-N)}.
\end{equation}
This reduces to $J_k(k)$ for integral $k$.  

We promote $k$ to an axionic field $k(x)$ and put it under the
integral sign in the CS action of Eq. (\ref{csaction}).  Take $k(x)$
to vanish outside some closed surface and to have a constant
non-integral value $k$ inside (except for some thin-wall transition
region). To follow L\"uscher, we consider the expectation value of
$\exp I_{CS}$ in a YM theory, which is  the same as $Z(k)/Z(0)$ of
YMCS theory. This result is given by replacing
$J_k(k)$ in Eq. (\ref{z3}) with $F_k(k)$ from Eq. (\ref{nonint}).
Because \cite{djt} the CS integral is a surface integral for the pure-gauge
configurations over which we are integrating, we have:
\begin{equation}
\label{surfint}
\langle e^{2\pi i kN_{CS}}\rangle_{YM} = 
\langle e^{2\pi ik\oint_S dS_iV_i}\rangle
\end{equation}
and:
\begin{equation}
\label{ymgauss}
\langle e^{2\pi i kN_{CS}}\rangle_{YM}=\frac{Z(k)}{Z(0)} = \exp
\{\frac{V_S}{V_c}e^{-I_c}[F(k)-1]\}
\end{equation}
where $V_S$ is the volume enclosed by the surface $S$ and $V_i$ is a
CS surface density (given explicitly for sphaleron-like configurations
in Eq. (\ref{surface}) below).  Because $F(k)\leq 1$, there is an
interpretation similar to L\"uscher's:  There is a bag, defined by the
surface where $k(x)$ changes, with an energy above the vacuum by an
amount proportional to the volume of the bag.  This bag is analogous
to the domain-wall sphaleron discussed in the next section.

Some qualitative information about the CS susceptibility can be
gleaned from the small-$k$ limit of Eq. (\ref{ymgauss}).  In this
limit,
\begin{equation}
\label{smallk}
F(k)\rightarrow 1-k^2(\frac{2\pi^2-9}{12})\equiv 1-\gamma k^2.
\end{equation}
 This form of the small-$k$ limit allows us to interpret the
 distribution of $N_{CS}$ as Gaussian:
\begin{equation}
\label{ymgauss2}
\langle e^{2\pi i kN_{CS}}\rangle_{YM}\rightarrow
\exp -\{2\pi^2k^2\langle N_{CS}^2\rangle \}.
\end{equation}      
Because we expect $\langle N_{CS}^2\rangle \sim V$, the Gaussian
expectation value vanishes in the infinite-volume limit.  In fact, by
comparing Eqs. (\ref{ymgauss}) and (\ref{ymgauss2}) we find an
approximate value for the CS susceptibility:
\begin{equation}
\label{cssus}
\frac{\langle N_{CS}^2\rangle}{V}=\frac{\gamma}{2\pi^2V_c}e^{-I_c}.
\end{equation}
This expression, while presumably not quantitatively accurate, is of a
form suggested earlier \cite{cy} in which the $d=3$ topological
susceptibility is of the form
\begin{equation}
\label{cyan}
\frac{\langle N_{CS}^2\rangle}{V}=\xi\langle \Theta \rangle
\end{equation}
where $\Theta$ is the trace of the stress-energy tensor and $\xi$
is a numerical constant.  For a dilute gas condensate,
\begin{equation}
\label{trace}
\frac{1}{3}\langle \Theta \rangle = \frac{1}{V_c}e^{-I_c}
\end{equation}
and $\xi=\gamma /6\pi^2$ from Eq. (\ref{cssus}). 

\section{Sphalerons and half-integral knots}
\label{knotsec}

A sphaleron has CS number $1/2$.  If sphalerons are dilute, they can be
idealized to  pure-gauge configurations.  These
configurations can be associated with fictitious Abelian field lines
through the Hopf fibration $S^3\rightarrow S^2$, with homotopy
$\Pi_3(S^2)\simeq Z$. The integer classes of this homotopy come from
an integral, the Hopf invariant, which is in fact the same as the
original CS number (see, {\it eg }, \cite{jahn,jp,co02}).  The Hopf
invariant $N_H$ is both a Gauss link number for the pre-images of any
two distinct points in $S^2$ in the Hopf fibration, and an
Abelian CS term for a fictitious Abelian gauge potential and
magnetic field.  Pre-images of $S^2$, necessarily closed curves, are
just field lines of this fictitious magnetic field, and so the Hopf
invariant expresses the linking of any two distinct closed field
lines. (For idealized Dirac-string center vortices
the CS number can also be expressed equivalently as a Gauss link
integral and as an Abelian CS term, but the normalization is
different, and the CS number can be half-integral in the simplest case.)

For the sphaleron the CS number is $1/2$; how can this be reconciled
with the link-number interpretation?  The answer is that in knot
theory \cite{knots} presented as two-dimensional graphs with over- and
under-crossings, each crossing contributes $\pm 1/2$ to the total link
number, just as does an isolated sphaleron.  In a certain sense, which
we make explicit below, sphalerons can be mapped onto these crossings.
Compact knots must have an even number of crossings; only knots
stretching to (and thus closed at) infinity  can have an odd number of
crossings in a  region excluding infinity.    So the sphaleron puzzle
comes down to how one closes the fictitious Abelian flux lines
that flow through the sphaleron.  We show here how this can
be done by introducing a domain-wall sphaleron containing the
other $1/2$ needed for integral CS number, and hence integral Hopf
invariant.  The domain wall can be, but is not required to be, on the
sphere at infinity.  If not, then the fictitious Abelian
field lines vanish identically outside the domain wall, which acts as
a superconducting wall for the fictitious field lines.

We give a second interpretation of the field-line knots,
which relates them to the formulation of $d=4$ topological charge as the
intersection of closed vortex and vortex-nexus surfaces.  This
interpretation maps the $d=4$ intersection numbers onto $d=2$
intersection numbers of closed lines (vortices) in the two-plane, some
of which must carry point nexuses and anti-nexuses.  In a formal
sense, the resulting formulation of half-integral CS number becomes a
two-dimensional projection of earlier formulas \cite{er00,co00} which
express $d=4$ $SU(2)$ topological charge as composed of components of
charge $\pm 1/2$, localized at the (assumed transverse) intersection
points of $d=4$ vortices and vortex-nexus combinations.  The total
(and integral) topological charge is computed as an intersection
integral with an extra weight factor coming from traces over the
Lie-algebra matrices of vortices and nexuses.  Both
interpretations will illustrate how an odd number of sphalerons requires
a sphaleron-like configurations at infinity.

\subsection{Sphalerons and link numbers of knots}
\label{sphlk}
The connection between the non-Abelian CS number of a pure-gauge
configuration $U$ and the Abelian linking number is found (see,
for example, \cite{jp,jahn,co00}) by exploiting the Hopf map
$S^3\rightarrow S^2$, with homotopy $\Pi_3(S^2)\simeq Z$, in the form
of a map from the $SU(2)$ group element to a unit vector $\hat{n}$:
\begin{equation}
\label{hopfmap}
U\tau_3U^{-1}\equiv \tau \cdot \hat{n}. 
\end{equation}
This is, of course, a compact map.
Since $U$ can be right-multiplied by  $\exp (i\alpha
\tau_3/2)$ without changing $\hat{n}$, each $\hat{n}$ corresponds to a
coset $SU(2)/U(1)$.  The linked curves in question are the pre-images
of points $\hat{n}$ on the sphere $S^2$.  This unit vector defines an
Abelian gauge potential and field, via:
\begin{equation}
\label{abela}
{\mathcal{A}}_i=iTr(\tau_3U\partial_iU^{-1});\
\end{equation}
\begin{equation}
\label{abelb}
{\mathcal{B}}_i=-i\epsilon_{ijk}Tr(\tau_3U\partial_jU^{-1}U\partial_kU^{-1})=
\frac{1}{2}\epsilon_{ijk}\epsilon_{abc}n^a\partial_jn^b\partial_kn^c.
\end{equation}
Because of the properties of the $\epsilon$ symbol and of group
traces, the non-Abelian CS integral of Eq. (\ref{ncs}) can be written
in terms of the Abelian field and potential:
\begin{equation}
\label{abelcs}
N_{CS}=\frac{1}{16\pi^2}\int d^3x {\mathcal{A}}_i{\mathcal{B}}_i\equiv N_H,
\end{equation}
where $N_H$ is the Hopf invariant, an integer characterizing the
homotopy class of the map.  The second equation in (\ref{abelb}) is
only true if Dirac strings are omitted.  For sphalerons no such
strings occur (see Eq. (\ref{abelsphb}) below).

The Hopf invariant is a link number of any two distinct field lines of
the field ${\mathcal{B}}_i$.  As  textbooks on knot theory discuss
\cite{knots}, these $d=3$ knots can be expressed in a
quasi-two-dimensional way, with graphs constructed
from over- and under-crossings of components of knots, and topological
invariance in $d=3$ reduced to Reidemeister moves in $d=2$.  (Another
good example of the $d=2$ nature of $d=3$ knots is Witten's derivation
\cite{wit89} of Jones polynomials from $d=2$ conformal field theory.)
In this picture, knot
components lie in a plane, except that they fail to intersect at an
over- or under-crossing by a vanishingly small distance
$\epsilon$.  Of course, in $d=3$ linked knot components may be very
far from touching one another, but in our case we are only interested in
 nearly-touching crossings,  so that contributions to the  Gauss link number are localized to these crossings.  The global $d=3$ topology
is not affected by this assumption.  
Then each crossing of distinct knot components
contributes an additive term $\pm 1/2$ to the conventional Gauss
linking integral, and there are no contributions from portions of the knot components 
separated by distances that are large compared to $\epsilon$.
(As we discuss below, this contribution of $\pm 1/2$ also
holds  for self-crossings of one component with itself, leading to
integral framed link number, because each self-crossing is actually a
double crossing.)  There is no contribution away from the crossings
even if the knot components extend in an arbitrary way (as long as components
do not cross each other) into all three dimensions. For closed compact
components there is always an even number of crossings and hence an
integral link number.  Half-integral linking numbers occur naturally
for non-compact knots, that is, knots with an odd number of
crossings.  For closed knot components, this can only occur
when the component curves are closed at infinity.
 
We give a specific example of these concepts.
A pure-gauge sphaleron centered at the origin is described by a gauge
function $U$ of the form:
\begin{equation}
\label{gaugesph}
U=\exp [i\beta( r)\vec{\tau}\cdot \hat{r}/2];\;\;\beta (0)=0;\;\;
\beta(\infty ) = \pi.
\end{equation}
One finds for the fictitious Abelian components:
\begin{equation}
\label{n}
\hat{n}=\hat{r}\cos \theta + \hat{\theta } \sin \theta \cos \beta +
\hat{\phi}\sin \theta \sin \beta;
\end{equation}
\begin{equation}
\label{abelsph}
{\mathcal{A}}_i =\hat{r}_i\beta^{\prime}\cos \theta +
\frac{\hat{\phi}_i}{\rho}(\cos \beta -1)\sin^2 \theta -
\frac{\hat{\theta }_i}{r} \sin \theta \sin \beta ;
\end{equation}
\begin{equation}
\label{abelsphb}
{\mathcal{B}}_i=\frac{2\hat{r}_i}{r^2}\cos \theta (\cos \beta
-1)+\frac{\hat{\phi}_i}{r}\beta^{\prime} (1- \cos \beta )\sin \theta +
\frac{\hat{\theta}_i}{r}\beta^{\prime}\sin \theta \sin \beta .
\end{equation}

These field lines have several important properties.  
First, the flux integrated over any sphere surrounding the origin is
zero, so there is no monopole and no Dirac string for this sphaleron.
Second, by inspection of Eq. (\ref{abelsphb}) one sees that there is
one, and only one, way in which the field lines can be terminated in a
finite region.  If for $r\geq a$, $\beta(r)$ takes the value
$2\pi N$ where $N$ is an integer, the field ${\mathcal{B}}_i$
vanishes identically for $r\geq a$. Since the fictitious field lines
are closed, this can only happen if the field lines run along the
surface of the sphere $r=a$ and at some point return to the vicinity
of the sphaleron and close.  Schematically, the field lines look like
those depicted in Fig. \ref{fig1}.  The bounding surface $r=a$ acts as
a superconductor for the Abelian field lines.  (Of course, this
bounding surface need not literally be a sphere, but it can be any
surface with the topology of $S^2$ that encloses the sphaleron.)
Once the field lines are compactified in this way, there is no problem
interpreting the Hopf invariant in terms of linkages of two of this
family of closed curves.  On the other hand, if $\beta (r)$ never
reaches $2\pi N$, it is easy to see from the explicit form of
Eq. (\ref{abelsphb}) that the field lines never return to the vicinity
of the sphaleron, but continue on to infinity.

A third important property is that in the vicinity of the sphaleron the field lines are sheared so that any two lines, projected into a plane, cross each other.  Below we will interpret this crossing as a contribution to the linkage of knotted field lines.

For the sphaleron $\beta (r)$ approaches $\pi$ asymptotically.  We
can, in analogy with the discussion above, bring the radius at
which $\beta = \pi$ to any desired finite value $r=b$, as long as $b$
is large compared to all natural length scales, such as $M^{-1}$.
This does not compactify it, because its Abelian field lines keep on
going past $r=b$.  But we can compactify it with a domain-wall
sphaleron at $r=a,\;a>b$, by increasing $\beta$ to $2\pi$ at $r=a$.
Then, as shown, the fictitious field lines close, and there is an
extra CS number of $1/2$ on the domain-wall sphaleron.

The CS number for the sphaleron can be found explicitly from the Hopf
invariant integral Eq. (\ref{abelcs}):
\begin{equation}
\label{cssph}
N_{CS}=-\frac{1}{2\pi }[\beta (\infty )-\sin \beta (\infty )]= -\frac{1}{2}.
\end{equation}
and of course it has the same value as would be obtained from the
spherical {\it ansatz} form of Eq. (\ref{spherncs}).  It can also be
written as a surface integral:
\begin{equation}
\label{surface}
N_{CS}=\int d^2S_iV_i;\;\;V_i=-\frac{\hat{r}_i}{8\pi^2r^2}
[\beta - \sin \beta ].
\end{equation}
Clearly, the contribution to $N_{CS}$ from the domain-wall sphaleron
can also be written as a surface integral over the domain wall.

\begin{figure}
\includegraphics{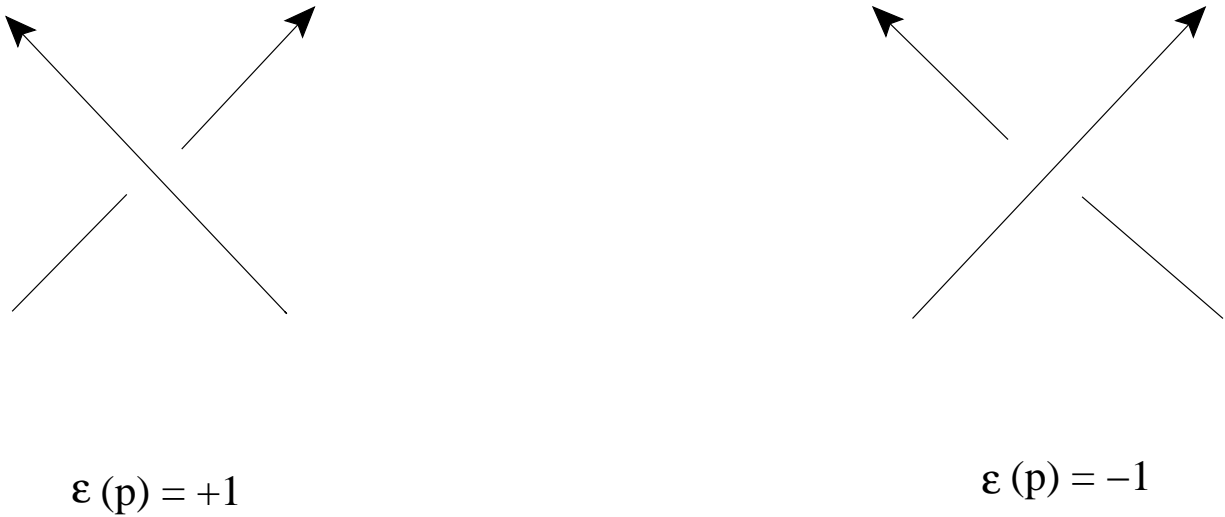}
\caption{\label{fig2} Over- and under-crossings and their values.}
\end{figure}

So what does a link number of $1/2$ mean for a sphaleron?  Recall
\cite{knots} how link numbers can be written as a sum of terms, each
of which is $\pm 1/2$.  The knots are displayed with suitable over- and
under-crossings in two-dimensional pictures.  For each crossing point
$p$ a factor $\epsilon (p)=\pm 1$ is defined as shown in Fig. \ref{fig2}.

For two distinct curves the link number $Lk$ is then defined as:
\begin{equation}
\label{lk}
Lk=\frac{1}{2}\sum_{p\in C}\epsilon (p)
\end{equation}
where $C$ is the set of crossing points of one curve with the other
(self-crossings will be discussed later).  This suggests that in some
sense a sphaleron is the topological equivalent of a single crossing,
with (as one quickly checks) an even number of crossings needed for
describing the linkage of closed compact curves.  Of course, since a
sphaleron is localized, one needs to interpret the crossings in
Figs. 1 or 2 as being infinitesimally separated.  This in itself is
not necessary for understanding the topology but it is necessary for
interpreting the topology in  terms of localized sphalerons.

We can express this in terms of the sort of integral occurring in the
formula (\ref{abellink}) for link number.  Consider the two infinite
straight lines
\begin{equation}
\label{stline}
z_i=(s\cos \alpha ,s \sin \alpha ,0);\;\;z^{\prime}_i = 
(t\cos \beta, t\sin \beta , \epsilon)
\end{equation}
where $-\infty < s, t < \infty$ with $ds,dt$ the elements of distance
along the lines.  Their distance of closest approach is $\epsilon$.
For a configuration of two infinite straight lines the value of
$\epsilon$ does not matter, but if the lines are part of a knot with
curvature,  $\epsilon$ must be treated as infinitesimal.  The
integrals in the formula
\begin{equation}
\label{stlink}
Lk= \frac{1}{4\pi}\int^{\infty}_{-\infty}ds
\int^{\infty}_{-\infty}dt
\epsilon_{ijk}\dot{z}_i\dot{z}^{\prime}_j
\frac{(z-z^{\prime})_k}{|z-z^{\prime}|^3}
\end{equation}
are readily done, and yield:
\begin{equation}
\label{lkvalue}
Lk= \frac{1}{2}{\rm sgn}
[\epsilon_{ijk}\dot{z}_i\dot{z}^{\prime}_j(z-z^{\prime})_k].
\end{equation}
In the course of evaluating the integral of Eq. (\ref{stlink}) in the
limit $\epsilon\rightarrow$0, one encounters standard definitions of
the Dirac delta function which allow one to write this integral for
the link number as:
\begin{equation}
\label{stlink2}
Lk= \frac{1}{2}\oint dz_i\oint dz^{\prime}_j\epsilon_{ij}\delta
(z-z^{\prime}){\rm sgn}\;  \epsilon
\end{equation}
where  ${\rm sgn}\; \epsilon$ refers to the sign of the
distance shown in Eq. (\ref{stline}) by which the two components are
separated out of the plane at their crossing points, that is, whether
there is an overcrossing or an undercrossing.

As long as one presents the knots as being quasi-two-dimensional,
which means that their components lie in one plane except for
infinitesimal displacements into the third dimension for crossings,
there are no other contributions to the integral for $Lk$, because the
triple product in its definition vanishes for curves lying in a plane.
As a result, in the present interpretation of link number, the link
number can be thought of as being localized, in units of $1/2$, to
points where the components of the knot appear to cross.  This is
quite similar to the interpretation in $d=4$ \cite{er00,co00,co02} of
$SU(2)$ topological charge as occurring in localized units of $1/2$.
The localization is associated with the intersection of surfaces
representing center vortices and vortex-nexus combinations, with an
analog in $d=2$ which we discuss below.

In fact, it is easy to see that away from the infinitesimally-close
crossing points, the knots may be arbitrarily deformed into the third
dimension as long as components do not cross each other, since the
difference of the contribution to $Lk$ from a $d=2$ component and one
deformed into $d=3$ is a Gauss integral with no linkages.  If, in this
process of deformation, other knot components become infinitesimally
close to each other, new contributions to the total $Lk$ of
$\pm 1/2$ will be generated, but their sum will be zero.

\subsection{Knots and $d=2$ intersection numbers}

The form of Eq. (\ref{stlink2}) for the link number is very
suggestive; aside from the sign function in the integrand and the
factor of $1/2$, it is the integral representation of the signed sum of
intersection numbers for curves lying in a plane.  In $d=4$,
the usual topological charge (the integral of $G\tilde{G}$)
for idealized pure-gauge $SU(2)$ center vortices and nexuses is also
represented by an intersection-number integral, including a factor of
$\pm 1/2$ coming from group traces \cite{er00,co00}.  The sign of this
group factor is governed by the presence or absence of nexuses and
anti-nexuses, each of which reverses the direction of the $SU(2)$
magnetic field lines lying in the vortex surface.  In $d=4$ center
vortices are described by closed two-surfaces, and nexus-vortex
combinations are described by such surfaces with a closed nexus world
line  lying in the vortex surface. For every nexus world line there is
an anti-nexus world line.  The intersection-number form can be
translated into a link-number form \cite{co00}, where the link is
between a center vortex with no nexus and a nexus (or anti-nexus)
world line. 
       
Here we give some simple examples of $d=3$ knot linkages represented
by $d=2$ graphs  which can be considered as the projection into two
dimensions of  $d=4$ vortex-nexus topological charge. There is no need
to distinguish over- and under-crossings; instead, the crossings are
interpreted as intersections of closed lines whose orientation changes
whenever a (point) nexus is crossed in the process of tracing out a
closed line.  The link number is calculated by counting (with signs)
the linkages of closed curves and nexuses or anti-nexuses.  In this
case a curve and a point are linked if the point is inside the curve;
otherwise they are unlinked.

\begin{figure}
\includegraphics{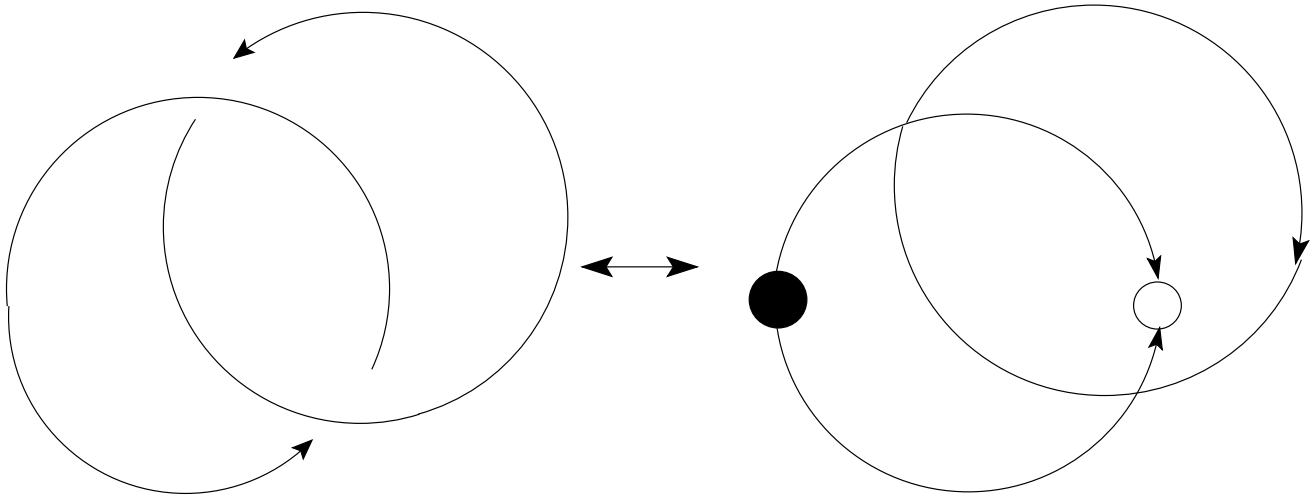}
\caption{\label{fig3}  A simple knot presented as a $d=2$ projection
of a center vortex and a center vortex with a nexus (filled circle)
and anti-nexus (open circle).}

\end{figure}

We obtain the $d=2$ CS integral, which is completely
analogous to the $d=4$ expression for center vortices and nexuses
\cite{co00}:
\begin{equation}
\label{d2cs}
N_{CS}=\sum_{\rm crossings}
\oint dz_i\oint dz^{\prime}_j\epsilon_{ij}\delta
(z-z^{\prime})Tr (QQ^{\prime})
\end{equation}
where $Q,Q^{\prime}$ take on the values $\pm \tau_3/2$, with the sign
depending on the orientation of segments of the closed curves.  The
orientation must change every time a nexus or anti-nexus is crossed in
the course of tracing out the curve.  Fig. \ref{fig3} illustrates this
formalism for a simple two-component knot represented both as an
over/undercrossing link and as a vortex-nexus link.  In the figure, a
filled-in circle is a nexus and an open circle is an anti-nexus; there
must be as many of one as of the other on any closed vortex curve. A
more detailed discussion of the correspondence between knots
and vortex-nexus ideas (including twist and writhe) will be given elsewhere.

In this way we can connect topological charges in dimensions
two, three, and four.  In all cases, for $SU(2)$ the localized unit of
topological charge is $\pm 1/2$, but compactification of the space under
consideration yields a sort of topological confinement of these
fractional units to integral totals.
 
\section{Linked and writhing center vortices}
\label{twist}

The standard center vortex \cite{co79} is an Abelian configuration,
essentially a Nielsen-Olesen vortex.  It contributes to CS number
through the $A\cdot B$ term, not through the $A^3$ term, and the
techniques used above to generate an Abelian potential and field are
irrelevant;  the vortex itself is Abelian, and in its
idealized pure-gauge version is described by  a closed Dirac-string
field line.  These closed lines may be linked, including the
self-linkages given in terms of twist and writhe.  Such
linkages generate CS number, as expressed through the $A\cdot B$
integral.  However, even integral link numbers give rise to CS numbers
whose quantum is $1/2$, while twist and writhe give rise to
arbitrary real CS number.

Generically, two distinct center vortices in $d=3$ never touch each
other, whether or not they are linked.  But to generate $d=4$
topological charge, which is a weighted intersection integral of the
points at which center vortices intersect (possibly with the
intervention of nexuses), two vortex surfaces must have common points.
If the intersection is transverse,
these points are isolated.  There must be a corresponding notion of
linked vortices touching each other in $d=3$ as well.  We can think
of vortices as closed strings in $d=3$, which evolve in
a Euclidean ``time'' variable (the fourth dimension). Vortices have
points in common at the isolated instant in which they change their
link number (reconnection) \cite{co95}.  To generate
topological charge, we must change the CS number, which is equivalent
to changing the link number.  Also
necessary in this simple situation is the presence of at least one
nexus, which reverses the sign of the vortex magnetic fields.  We
discuss some elementary cases in which reconnection changes the CS
number by $\pm 1/2$, and in which, if compactness in $d=4$ is demanded,
the overall change in link number yields integral topological charge.
Note that compactness in $d=3$ has nothing to do with compactness in $d=4$
(consider the product $S^3\times R$). Even for reconnection that
changes the writhe of a single vortex, which can be arbitrary, it is
possible to have changes in CS number quantized in units of $1/2$.  The
appearance of this unit of $1/2$, plus the pairing of $d=4$ intersection
points of compact surfaces \cite{er00,co00,co02} is somewhat analogous
to the pairing of over- and under-crossings for compact $d=3$ knots,
discussed above.

In considering the evolution in time of various field configurations
carrying topological charge,  note that there are real differences
between $d=4$ topological charge interpolated by sphalerons and by
reconnection of vortices.  A sphaleron is the (unstable) saddlepoint
of a {\em classical} path in configuration space.  One can extend the
sphaleron gauge angle $\beta (r)$ to a function $\beta (r,t)$
with $\beta(r,0) \equiv \pi$ and
which yields unit topological charge in the form $(1/(2\pi )[\beta
(r,\infty )-\beta (r,-\infty )]=1$,
such as $\beta =2\arctan (r/t)$.  There is no need to pair the
sphaleron with another sphaleron.  A vortex, however, cannot evolve
classically since it must reconnect and overlap with itself or with
another vortex.  The action penalty from overlap yields a tunneling
barrier, and thus reconnections with half-integral CS number
must be paired.

It would take another paper to discuss all the ramifications of vortex
self-linkage, including the role of nexuses, and the new twisted nexus
presented recently \cite{co02}.  We restrict ourselves here to a few
simple examples, including a new  Abelian twisted vortex, and some
general conclusions.  The main point is that self-linking, whether
considered for idealized Dirac-string vortices or for fat physical
vortices, leads to contributions to $N_{CS}$ that can be
essentially arbitrary real numbers, although the self-linking is
spatially localized.  As for sphalerons, one can introduce a domain
wall to carry extra CS number, bringing the total to an integer.

\subsection{Linking of distinct vortices and half-integral CS number}
\label{distvort}

For pure-gauge center vortices the interpretation of $N_{CS}$
in terms of a link number is straightforward, if two distinct
vortices are linked, but more troublesome if self-linking
 is involved.  For the straightforward case of
linking of distinct vortices the CS number is half the link number and
can therefore be half-integral.  If it is half-integral, the
configurations is not compact, even though the links composing the
two vortices are spatially compact.  If these links have maximum
spatial scale $L$, the gauge potential from the vortices behaves as
$L^2/r^3$ when $r\gg L$, and so it falls off sufficiently
rapidly at large distances that no surface terms arise in various
integrals of interest.

The gauge parts of  two distinct center vortices are described by
closed curves $\Gamma,\Gamma^{\prime}$:
\begin{equation}
\label{abelvortexa}
A(x;\;\Gamma )_i=(\frac{2\pi\tau_3}{2i})
\epsilon_{ijk}\partial_j\oint_{\Gamma} dz_k\Delta (x-z);
\end{equation}  
\begin{equation}
\label{abelvortexb}
B(x;\;\Gamma^{\prime} )_i=(\frac{2\pi\tau_3}{2i})
\oint_{\Gamma^{\prime}} dz_i\delta (x-z)
\end{equation}
where $\Delta (x-z)$ is the free massless propagator in $d=3$.  The CS
number of the mutual linkage of $\Gamma ,\Gamma^{\prime}$ is:
\begin{equation}
\label{abellink}
N_{CS}=\int d^3xTrA(x;\;\Gamma )_iB(x;\;\Gamma^{\prime} )_i
=(\frac{-1}{2})Lk(\Gamma,\Gamma^{\prime});
\end{equation}
\begin{equation}
\label{deflk}
Lk(\Gamma,\Gamma^{\prime})\equiv \oint dz_i
\oint dz^{\prime}_j\epsilon_{ijk}\partial_k\Delta (z-z^{\prime}).
\end{equation}

\begin{figure}
\includegraphics{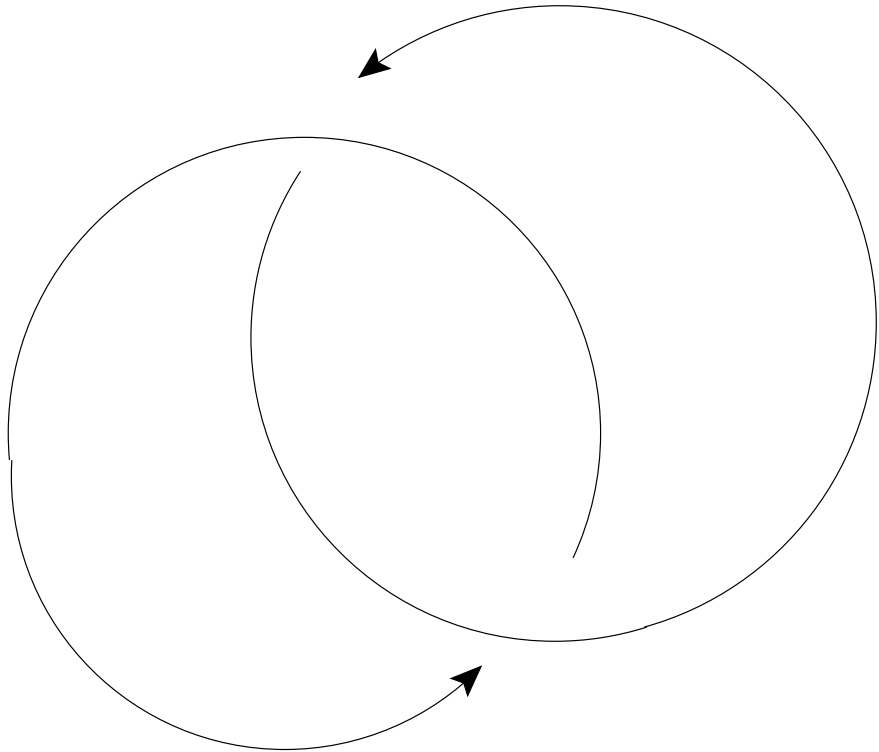}
\caption{\label{fig4} A simple two-component knot with no twist or writhe.}
\end{figure}

If curves $\Gamma ,\Gamma^{\prime}$ are linked, as in Fig. \ref{fig4},
the corresponding CS number is $1/2$,  because of the factor $1/2$ in
front of the $Lk$ integral in Eq. (\ref{abellink}).

\subsection{Self-linkage of Dirac-string vortices}

\begin{figure}
\includegraphics{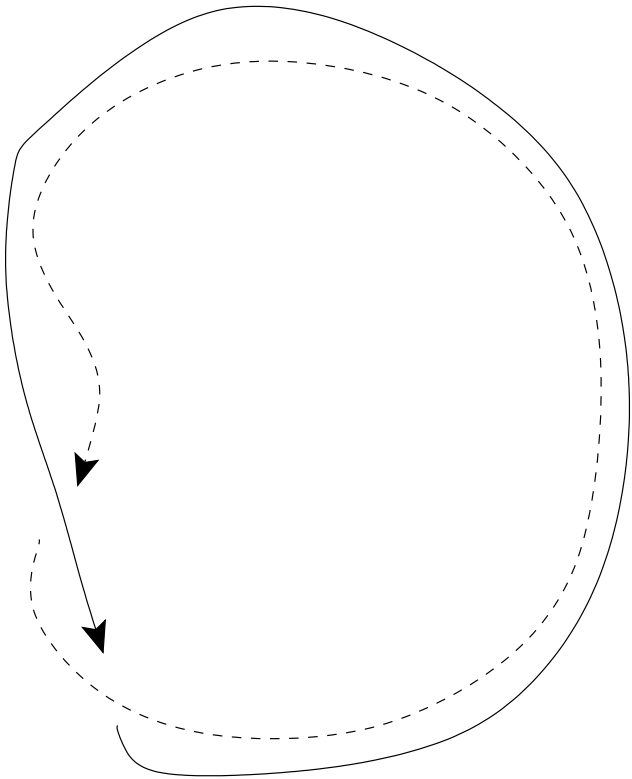}
\caption{\label{fig5} Ribbon framing of a single-component knot with
one unit of twist and two crossings.}

\end{figure}

Self-crossings of a single vortex Dirac string give rise to twist
($Tw$) or writhe ($Wr$).    With the usual sort of ribbon framing
\cite{knots} used to define self-crossings, neither twist nor writhe
is a topological invariant and neither is restricted to integral
values.   Their sum,  the framed link number $FLk$, is an
integer-valued topological invariant whose value depends on the
framing.  A simple ribbon-framing is shown in
Fig. \ref{fig5}.  The CS number is not the integer $FLk$;
instead, it is the writhe $Wr$, or self-link integral, given in
Eq. (\ref{lkvalue2}) below.

Because the vortex is Abelian, $N_{CS}$ receives contributions only
from the $\vec{A}\cdot \vec{B}$ term \cite{co95,cy}:
\begin{equation}
\label{twistcs}
N_{CS}=\frac{-1}{8\pi^2}\int d^3x Tr\vec{A}\cdot
\vec{B}=\frac{1}{4}Lk(\Gamma ,\Gamma)
\end{equation}
where $Lk(\Gamma ,\Gamma )\equiv Wr$ is the self-linking number or
writhe of Eq. (\ref{lkvalue2}). The writhe can be anything, depending
on the geometry of the vortex.

For Frenet-Serret framing (displacing the ribbon infinitesimally from
the curve $\Gamma$ along the principal normal vector $\hat{e}_2$) the
twist is:
\begin{equation}
\label{twdef}
Tw=\frac{1}{2\pi}\oint ds \hat{e}_2\cdot \frac {d\hat{e}_3}{ds}
\end{equation}
where $\hat{e}_3$ is the binormal vector.  It too is 
geometry-dependent and not restricted to be an integer or
simple fraction.

A typical self-crossing is shown in Fig. \ref{fig1}, which was
introduced to illustrate a center vortex.  We now interpret that
figure as a picture of twisting but unwrithed fictitious Abelian field
lines (the discussion is essentially the same if one replaces ``twist''
by ``writhe''; the two are interconvertible).  Even though this is a
compact knot, it appears that there is only one crossing. Actually
there are two for the framed knot of Fig. \ref{fig5}. For an untwisted
curve the  Gauss link number for the writhing curve $\Gamma$ is equal
to the writhe:

\begin{equation}
\label{lkvalue2}
FLk(\Gamma ,\Gamma )=Wr=\frac{1}{4\pi}\oint_{\Gamma}
dz_i\oint_{\Gamma}
dz^{\prime}_j\epsilon_{ijk}\frac{(z-z^{\prime})_k}{|z-z^{\prime}|^3}.
\end{equation}
As the contours are traced out the crossing point is encountered
twice, so the value of $FLk$ in Eq. (\ref{lkvalue2}) is $(1/2)+(1/2)=1$.
Or one may calculate $FLk$ by counting the crossings of the link with
its ribbon frame; again there are two crossings.

Note that the same value of the writhe applies to the center vortex of
Fig. \ref{fig1}, but because of group traces the CS number is, for
gauge group $SU(2)$, half the writhe.  We see that topologically a
unit of writhe in the fictitious Abelian field lines corresponds to
two sphalerons, but a unit of writhe in a center vortex corresponds to
only one sphaleron.

We note  that the self-linking number of a pure-gauge center vortex,
as described in Eq. (\ref{abelvortexa}), is also the self-flux of the
corresponding Abelian potential of Eq. (\ref{abela}):
\begin{equation}
\label{selfflux}
Lk(\Gamma ,\Gamma )=Wr=\frac{1}{8\pi}\oint_{\Gamma} dz_i{\mathcal{A}}_i(z).
\end{equation}

The simplest case of dynamic reconnection of a center vortex begins
with a configuration such as that shown in Fig. \ref{fig1}, to which we
assign some twist $Tw$ and writhe $Wr$, whose sum is an integer, the
framed link number.   Reconnection,
which changes the overcrossing shown in the figure
to an undercrossing, changes the framed link number by 2, not 1, as
one can appreciate from a study of Fig. \ref{fig5}. If we assume
that the lines shown in Fig. \ref{fig1} are 
separated by an infinitesimal distance $\epsilon$ at crossing, then
the twist, which is a purely geometric quantity, will change only by
$O(\epsilon)$.   The upshot is that the writhe
changes by 2 and the CS number changes by $1/2$, because of the factor
of 1/4 in Eq. (\ref{twistcs}).  So certain cases of writhe
reconnection lead to a quantum of $1/2$ for $N_{CS}$, just as for simple
mutual linkages. As discussed above, compactness in $d=4$ requires these
acts of reconnection to be paired, leading to integral topological
charge but quantized in units of $1/2$.

Let us conclude this subsection with a new and simple special case of
a twisting vortex with half-integral CS number.  This vortex is
Abelian, described  by the gauge function
\begin{equation}
\label{twvortex}
U=\exp \{\frac{i\tau_3}{2}[\phi+\gamma (z)]\};\;\;A_i=U\partial_iU^{-1}=
(\frac{\tau_3}{2i})[\frac{\hat{\phi}_i}{\rho}+\hat{z}_i\gamma^{\prime}(z)].
\end{equation}
Here $\phi,z$ are the usual cylindrical coordinates.
The magnetic field comes from the Dirac string in the vector potential:
\begin{equation}
\label{twistb}
B_i(x)=(\frac{2\pi \tau_3}{2i})\hat{z}_i\delta (x)\delta (y).
\end{equation}
Evidently this vortex lies along the $z$-axis.  In order to describe a
vortex which is a closed loop of length $L$, we should identify $z=0$
with $z=L$.  This requires that the gauge function $U$ be the same at
these two values of $z$, or that
\begin{equation}
\label{gammaeq}
\gamma (L)-\gamma (0)=4\pi N
\end{equation}
for some integer $N$.  The integral in Eq. (\ref{twistcs}) is trivial,
and yields
\begin{equation}
\label{twcs}
N_{CS}=\frac{1}{8\pi}[ \gamma (L)-\gamma (0) ]=\frac{N}{2}.
\end{equation}
So such a twist is equivalent to $N$ sphalerons.

Not unexpectedly, one can get any desired value for the CS number by
decompactifying; that simply removes the requirement in
Eq. (\ref{gammaeq}) on the difference of $\gamma$ at the endpoints.

\subsection{Writhe and collective coordinates for fat vortices}

We next move from idealized Dirac-string vortices to
physical vortices, composed of flux tubes whose thickness is
essentially $M^{-1}$.  There is not only the YM vortex described in
\cite{co79}, but also in YMCS theory there is \cite{co96} a
self-conjugate center vortex.  The example of \cite{co96} has no twist
or writhe, and it has a purely imaginary CS action and therefore no CS
number.  But if this vortex (or the YM vortex) is
twisted, it will yield a contribution to the CS number which
is not constrained to be an integer or any simple fraction   This is
familiar in magnetohydrodynamics \cite{spit}, where CS number becomes magnetic
helicity.  The helicity is closely related to the so-called
rotational transform, or average angular displacement of a magnetic
field line per turn, in a plasma device such as a stellerator
or tokamak; this too is unconstrained.
 
For a physical center vortex, it was shown some years ago \cite{co95}
that center vortices arising from the YM action with a dynamical mass
term as in Eq. (\ref{dynmass}) lead to the replacement of
Eq. (\ref{twistcs}) by:
\begin{equation}
\label{fatcs}
N_{CS}=(\frac{1}{4})\frac{1}{4\pi}
\oint_{\Gamma} dz_i \oint_{\Gamma} dz^{\prime}_j\epsilon_{ijk}
\frac{(z-z^{\prime})_k}{|z-z^{\prime}|^3}F(R)
\end{equation}
where $R=|z-z^{\prime}|$ and
\begin{equation}
\label{fr}
F(R)=\frac{1}{2}\int_0^{MR}dvv^2e^{-v}.
\end{equation}   
For $MR\rightarrow \infty$, $F(R)\rightarrow 1$ and one recovers the
usual writhe integral, but for $MR\rightarrow 0$, $F(R)\simeq
(MR)^3/6$.  Because of this benign short-distance behavior,
ribbon-framing is irrelevant and there is no good distinction between
twist and writhe.  Clearly, the simple dynamical mass term of
Eq. (\ref{dynmass}) is at best a drastic simplification of complicated
quantum corrections leading to a dynamical mass, and whatever the real
form of the true $N_{CS}$ is, it will be a functional of various
collective coordinates describing the physical center vortex.

We can give a speculative and simplistic description of this
collective coordinate.  Whatever the true CS number of a vortex is, it
can be reduced to an integer (or more generally a rational fraction,
such as $1/2$ or $1/4$) by a non-compact gauge transformation.  This gauge
transformation is described in Eq. (\ref{gaugesph}), and is
characterized by an angle $\beta (r)$.  The value of $\beta (\infty )$
for this gauge transformation is determined by the original CS number
of the vortex, and can be treated as a stand-in for the collective
coordinates of this vortex.  The set of values of  $\beta (\infty )$
for the vortex condensate can then be integrated over, as we did for
sphalerons, and with the same effect:  Dynamical compactification and
CS number carried on domain walls.
 
\section{Fermions}
\label{fermions}

One way to obtain a CS term in $d=3$ is to start from ordinary
YM theory and integrate out a fermion doublet \cite{nise,anr}.  Thus we expect
that the same effects we have seen in YMCS theory should also be
visible as effects of fermions coupled to gauge fields.  In this
section we will make this connection concrete, and see how the effects
of the CS term emerge explicitly in terms of fermions.

It is well-known  that fermions or their solitonic equivalents
skyrmions can have exotic fermion number $F$ \cite{gw,gj}, and that
interactions of fields with gauge fields in the presence of a CS term
can lead to exotic statistics \cite{wz}.  In condensed-matter physics,
half-integral spin leads to half-integral CS level \cite{amp,dpw}, and
the CS term turns bosons into fermions.  Fermion zero modes bound by
solitons lead to puzzles about apparent fractional fermion number and
violation of BPS bounds in supersymmetry
\cite{super1d,SUNY,svv}.  For $d=3$ YMCS the resolution of such
puzzles will involve fermion zero modes at infinity which converts
local fractional fermion number to a global integer. This is the zero
mode associated with the domain-wall sphaleron at infinity.

An $SU(2)$ theory with an odd number of two-component fermions is
inconsistent because of the non-perturbative Witten anomaly in $d=4$
\cite{wit82}.  In $d=3$ an odd number of two-component fermions leads
to an odd CS level $k$ and dynamical compactification. 

\subsection{Zero modes and fermion number $1/2$}

In $3+1$ dimensions, the sphaleron sits halfway between vacua
differing by unit CS number.  A path between these vacua
correspondingly has unit anomalous violation of fermion number and
therefore the sphaleron carries fermion number $F=1/2$.  In general, 
the fermion number of a soliton background can be calculated in terms
of the asymmetry of the fermion spectrum.  The sphaleron is symmetric
under simultaneous rotations in physical space and isospin space, so
that grand spin $\vec G = \vec J + \vec I$ is conserved.  We can thus
decompose the solutions to the Dirac equation into channels labeled
by grand spin $G$.  In each channel with $G\neq 0$, we obtain an
eight-component spinor (describing the spin and isospin), describing
four distinct degrees of freedom.  In $G=0$, we have a four-component
spinor, describing two distinct degrees of freedom.  In both cases,
these spinors have the usual degeneracy factor of $2G+1$, and we write
the total fermion number as a sum over channels:  $F=\sum_G (2G+1)F_G$.

In each channel, the density of states in the continuum is related to
the total phase shift $\delta_G(\omega)$ by \cite{BB,fgjw}
\begin{equation}
\rho_G(\omega) = \frac{1}{\pi} \frac{d\delta_G(\omega)}{d\omega}
\end{equation}
so that integrating over the energy and including the contribution of
the bound states, we obtain the fermion number:
\begin{equation}
F_G = \frac{1}{2\pi} \left( \delta_G(m) - \delta_G(\infty) - \pi n_G^+
+ \pi n_G^-  - \delta_G(-m) + \delta_G(-\infty) \right)
\label{lev}
\end{equation}
where $n_G^+$ and $n_G^-$ are the number of positive- and
negative-energy bound states respectively.  We can obtain arbitrary
fractional values \cite{gw} for the fermion number from the phase
shift at infinity, which is sensitive only to the topological
properties of the background field.  It appears from this formula that a
$CP$-invariant configuration such as the sphaleron cannot carry net
fermion number, since the spectrum is symmetric in $\omega\to
-\omega$.  But there is a loophole:  the sphaleron has a single zero
mode, which will produce a fermion number of $\pm 1/2$, with the sign
depending on whether we include the zero mode with the positive or
negative energy spectrum \cite{JR}.  Just as we saw with link number
in Section \ref{knotsec}, the fermion number in Eq.~(\ref{lev}) is
generally an integer, but it is really a sum  of half-integral pieces,
and the sphaleron represents an exceptional case in which one of these
half-integers is not paired.  We will see that the extra zero mode
lives at infinity, in agreement with the knot-theoretic picture.

We will want to focus on the zero mode solutions to the Dirac
equation, which will occur only in the $G=0$ channel.  In this
channel, the Dirac equation reduces to an effective one-dimensional
problem, so we start by reviewing the properties of soliton zero modes
in $1+1$ dimensions.

\subsection{Zero modes in $1+1$ dimensions}

The simplest example of a soliton with fermion number $1/2$ is the
kink in $1+1$ dimensions \cite{JR}.  The Dirac equation is:
\begin{equation}
\gamma^0 \left(-i\gamma^1 \frac{d}{dx} + m \phi_1(x) \right)\psi(x) 
= \omega \psi(x)
\label{Diraceq}
\end{equation}
where we will work in the basis $\gamma^0 = \sigma_2$, $\gamma^1 = i
\sigma_3$ for the two-component spinor $\psi$.  In this section, $m$
is the fermion mass and not the CS mass.  The scalar background
$\phi_1(x)$ goes from $-1$ at $x=-\infty$ to $+1$ at $x=+\infty$, and we
will assume that $\phi_1(x) = -\phi_1(-x)$.  The detailed shape of
$\phi_1(x)$ will not be important for this discussion.  Just from the
topology, we see that we have a zero mode:
\begin{equation}
\psi_0(x) = \pmatrix{ e^{-m\int_0^x \phi_1(x')dx'} \cr 0} \,.
\label{zero}
\end{equation}

All nonzero eigenvalues of eq.~(\ref{Diraceq}) occur in
complex-conjugate pairs.  From a spinor $\psi_\omega(x)$ with eigenvalue
$\omega$, the the solution with eigenvalue $-\omega$ is
$\psi_{-\omega}(x) = \gamma_5 \psi_\omega(-x)$ where in our
basis, $\gamma_5 = \sigma_1$.  For the zero mode, however, we obtain:
\begin{equation}
\psi_{1}(x) = \pmatrix{ 0\cr e^{m\int_0^x \phi_1(x')dx'} }
\label{nonzero}
\end{equation}
which is non-normalizable.  This mismatch, which does not occur for
the analogous bosonic problem, is responsible for the nonzero quantum
correction to the mass of the supersymmetric kink \cite{super1d}.
It is also the underlying reason for the appearance of half-integer
fermion number, since all the other contributions to Eq.~(\ref{lev})
cancel between positive and negative energies.  The result is a
fermion number of $\pm 1/2$, with the sign depending on whether we
count the zero mode as a positive- or negative-energy bound state.
To lift this ambiguity, we could introduce a small constant
pseudoscalar field with interaction $\bar\psi i\gamma_5 \phi_2 \psi$,
which breaks the symmetry of the spectrum.  For $\phi_2$ small, the 
effect of this field is just to  change the energy of the zero mode
slightly (with the direction depending on the sign of $\phi_2$), which
fixes the sign precisely.  We will discuss this case further below.

For later reference, we note that we can characterize the normalizable
and non-normalizable solutions in a basis-independent way using
\begin{equation}
\gamma^1 \psi_0(x) = i\psi_0(x)
\end{equation}
for the normalizable zero mode while
\begin{equation}
\gamma^1 \psi_{1}(x) = -i\psi_{1}(x)
\label{bagboundary}
\end{equation}
for the non-normalizable mode.  (For an antisoliton, the situation is
reversed.)  So far we have just considered the localized effects near
a single kink, using scattering boundary conditions.  But in a
physical system, we also have to consider what is happening at the
boundary \cite{svv,SUNY}.  We can either place an antisoliton very far
away, so that the boundary can be made periodic, or we can put the
soliton in a box.  Both have the same effect, which is to allow the
other zero mode, Eq.~(\ref{nonzero}), to become a normalizable state
living far away.  In the former case, it is a zero mode localized at the
antisoliton.  For finite separation, both modes are displaced slightly
from zero by equal and opposite amounts, giving a symmetric spectrum.
In the latter case, the other zero mode lives at the walls; the
condition in Eq.~(\ref{bagboundary}) becomes simply a bag boundary
condition at the walls.

\subsection{Sphaleron zero modes}

The $d=3+1$ sphaleron case is closely analogous to the 1+1 case;
indeed, the spherical {\it ansatz} of Eq. (\ref{sphera}), with
fermions obeying Eq. (\ref{gspin}) below, maps directly on to $d=1+1$
fields.  We use the same notation as in Eq. (\ref{sphera}), with a
scalar field $\phi_1$, a pseudoscalar field $\phi_2$, and the space
component $H_1$ of an Abelian gauge potential $H_{\mu}$ (the time
component $H_0$ is zero for static configurations), and consider only
s-wave fermions.  Since the fermionic interactions induce an
effective CS term, we do not need to introduce one explicitly.
Following earlier work
\cite{dhn,ct,Yaffe}, we consider a fermion in the presence of a
sphaleron background in  the spherical ansatz.  In the grand spin
channel $\vec{G}=0$, where $\vec{G}=\vec{L}+(1/2)[\vec{\sigma}+\vec{\tau}]$ and
for s-waves $\vec{L}=0$, we have the fermion wavefunction:
\begin{equation}
\psi_L(x) = \left(f(r,t) + ig(r,t)\right) \Xi
\end{equation}
where $\Xi$ is a constant spinor with
\begin{equation}
\label{gspin}
\left(\vec \sigma + \vec \tau\right) \Xi = 0
\end{equation}
and $\vec \sigma$ and $\vec \tau$ are Pauli matrices
corresponding to spin and isospin respectively.  We normalize $\Xi$
so that $\Xi^\dagger \Xi=1$. Defining
\begin{equation}
\psi(r,t) = r \pmatrix{f(r,t) \cr g(r,t)}
\end{equation}
we find that the two-component spinor $\psi(r)$ obeys the
one-dimensional Dirac equation:
\begin{equation}
\left( i \gamma^\mu D_\mu + \frac{1}{r}(\phi_1(r) +
i\gamma_5 \phi_2(r)) \right) \psi(r,t) = 0
\end{equation}
where $\mu=0,1$ and $D_\mu = \partial_\mu + i H_\mu \gamma_5/2$ is the
covariant derivative for the 1+1 Abelian gauge potential $H_{\mu}$.

We can use the $U(1)$ symmetry of the spherical ansatz
to choose our gauge so that $H_1=0$ and, as stated above, for a stationary
configuration we will have $H_0=0$ as well.  Thus we can take
$\psi(r,t) = e^{i\omega t} \psi(r)$.  Since the sphaleron is
$CP$-invariant, the Higgs field $\bar\phi_1$ that we obtain must be
real, so the phase angle $\beta$  of Eq. (\ref{sphera}) must be an
integral multiple of $\pi$. For $\beta = \pi$ the Dirac equation
becomes:
\begin{equation}
\gamma^0 \left( -i \gamma^1 \frac{d}{dr} - \frac{\bar\phi_1}{r}
\right) \psi(r) = \omega \psi(r)
\end{equation}
which has the normalizable solution
\begin{equation}
\psi_0(r) = e^{+\int_{r_0}^r dr' \frac{\bar\phi_1(r')}{r'}}\psi_0(r_0)
\end{equation}
where $\gamma^1 \psi_0(r_0) = i\psi_0(r_0)$.  The situation is now exactly
analogous to the case of the kink:   We also have a corresponding
non-normalizable mode, given by
\begin{equation}
\label{newzm}
\psi_1(r) = e^{-\int_{r_0}^r dr' \frac{\bar\phi_1(r')}{r'}}\psi_1(r_0)
\end{equation}
with  $\gamma^1 \psi_1(r_0) = -i\psi_1(r_0)$.  As with
the kink, we can construct a pair of sphalerons with integer fermion
number and an even number of (near-)zero modes \cite{ct}.

\subsection{Level crossing and dynamical compactification}

We found the sphaleron zero mode as a normalizable solution to the
time-independent Dirac equation in $3+1$ dimensions with eigenvalue
zero.  We can then consider these three dimensions by themselves as a $d=3$
Euclidean spacetime.  The zero mode represents a level crossing
in the instantaneous eigenvalue of the $2$-dimensional Dirac equation
evaluated as a function of the $d=3$ Euclidean time variable $\tau$
\cite{anr}.  We can use this level crossing picture to understand
the dynamical action penalty for noncompact configurations.

Since the zero mode has $\vec{G}=0$ (see Eq. (\ref{gspin})), the level
that crosses zero must have equal and opposite spin and isospin.
Reducing to a $d=3$ theory, however, where we used to have a
four-component Dirac equation for each isospin component, we can now
consider just a two-component spinor, since the spin up and down
states can no longer be rotated into one another.  Thus we can have,
for example, a sphaleron background in which a spin-up isospin-down
state crosses from below to above, creating a fermion.  The
corresponding crossing in the other direction, which would create an
antifermion that could annihilate with this fermion through
gauge-boson exchange, is not normalizable.  Thus if this sphaleron is
not paired with a compensating antisphaleron, we will pay an action
penalty for this fermion proportional to the Euclidean time extent of
the system.

\subsection{Fermion number and Chern-Simons number}

  Although the CS term induced by fermions is just one term in
the derivative expansion of the fermion determinant, it gives the
entire contribution to the phase of the fermion determinant.  In the
language of the three-dimensional Dirac equation, the CS term is
simply the fermion number, which can be shown directly from the
effective action \cite{nise}, where it emerges as a result of the
$1+1$ dimensional chiral anomaly, or by explicitly considering the
contribution of each mode \cite{wit89}.  These works relate the
Chern-Simons number to the ``eta invariant'':
\begin{eqnarray}
N_{CS} = \eta \equiv -\frac{1}{2} \lim_{s\to 0} \sum_i {\rm sgn} \,
\omega_i |\omega_i|^{-s}
\label{etainvariant}
\end{eqnarray}
which in the continuum becomes:
\begin{equation}
\eta = F \equiv \sum_G (2 G + 1) F_G
\end{equation}
where $F_G$ is computed from Eq.~(\ref{lev}) with appropriate
regularization \cite{fgjw}.

The first paper of Ref.~\cite{wit89} gives a particularly simple
explanantion for the emergence of the eta invariant as the phase of
the determinant:  For a bosonic theory, each mode in the determinant
contributes
\begin{equation}
I_j = \int_{-\infty}^{\infty} \frac{dx_j}{\sqrt{\pi}}
e^{i\omega_j x_j^2}
= \lim_{\epsilon\to 0} \int_{-\infty}^{\infty}
\frac{dx_j}{\sqrt{\pi}} e^{i\omega_j x_j^2} e^{-\epsilon x^2}
= \frac{1}{|\sqrt{\omega_j}|} e^{i\frac{\pi}{4} {\rm sgn} \, \omega_j}
\end{equation}
and correspondingly for fermions we have:
\begin{equation}
I_j = |{\sqrt{\omega_j}}| e^{-i\frac{\pi}{4} {\rm sgn} \, \omega_j}
\end{equation}
leading to a total phase given by to Eq.~(\ref{etainvariant}).

We have seen above how the existence of half-odd-integral fermion number is
intimately connected to the boundary properties of the theory.  If we
allow background field configurations with more general boundary conditions,
violating both $C$ and $CP$, we can obtain an arbitrary fractional
fermion number, which still depends only on the topological properties
of the background field at infinity \cite{gw}.  These fractions will
enter the phase shift representation of the fermion number through
$\delta_G(\pm\infty)$.

Again, we will start by considering a one-dimensional example, which
will carry over directly to the $G=0$ channel in three dimensions.
We consider the Dirac equation:
\begin{equation}
\gamma^0 \left(-i\gamma^1 \frac{d}{dx} +
m (\phi_1(x) + i \gamma_5 \phi_2(x)) \right)\psi = \omega \psi
\label{Diraceq5}
\end{equation}
where we have introduced the pseudoscalar field $\phi_2(x)$.  For
concreteness, we will consider a definite background field
configuration, though as before the results do not actually depend on
the details of the field configuration, only its topology.  We take
the background that was considered in \cite{Dunne}:
\begin{eqnarray}
m\phi_1(x) &=& \mu \tanh \frac{\mu x}{2} \cr
m\phi_2(x) &=& \nu
\end{eqnarray}
where $m^2 = \nu^2 + \mu^2$.  To simplify the calculation, we have
chosen a reflectionless background, but the results we obtain are
generic.  The Dirac equation is now:
\begin{equation}
\pmatrix{ \nu & i\left( \frac{d}{dx} - \mu \tanh \frac{\mu x}{2}\right) \cr
i\left( \frac{d}{dx} + \mu \tanh \frac{\mu x}{2}\right) & -\nu }
\pmatrix{\eta_2(x) \cr c_\omega \eta_1(x)}
= \omega \pmatrix{\eta_2(x) \cr c_\omega \eta_1(x)}
\end{equation}
with $c_\omega = {\rm sgn} \,(\omega) \sqrt{\frac{\omega - \nu}{\omega
+ \nu}}$.  Squaring this equation, we find that the wavefunctions are solutions
to the Schr\"odinger equation for potentials of the reflectionless
P\"oschl-Teller form (see for example \cite{super1d} and references therein),
\begin{eqnarray}
\left( -\frac{d^2}{dx^2} - \frac{\mu^2}{2} {\rm sech}^2 \, \frac{\mu x}{2}
\right) \eta_1(x) &=& k^2 \eta_1(x) \cr
\left( -\frac{d^2}{dx^2} - \frac{3\mu^2}{2} {\rm sech}^2 \, \frac{\mu x}{2}
\right) \eta_2(x)&=& k^2 \eta_2(x),
\label{PT}
\end{eqnarray}
where $k=\sqrt{\omega^2 - m^2}$.

An incoming wave from the left is given by:
\begin{equation}
\psi_{\rm in}(x) = \pmatrix{1 \cr \lambda}  e^{ikx}
\end{equation}
where $\lambda = - \frac{k+i\mu}{\omega + \nu}$.  Propagating this
solution through the potential, the transmitted wave is:
\begin{equation}
\psi_{\rm out}(x) = \pmatrix{e^{i\delta_2(k)} \cr \lambda
e^{i\delta_1(k)}  e^{ikx}}
\end{equation}
where
\begin{eqnarray}
\delta_1(k) &=& 2 \arctan \frac{\mu}{2k} \cr
\delta_2(k) &=& \delta_1(k) + 2 \arctan \frac{\mu}{k}
\end{eqnarray}
are the phase shifts of the reflectionless Schr\"odinger equations in
(\ref{PT}).  To compute the fermion phase shift, we compare $\psi_{\rm
out}$ to the spinor $\psi_{\rm rot}$ obtained by performing the chiral
rotation on $\psi_{\rm in}$ that rotates it from the vacuum on the
left to the vacuum on the right,
\begin{equation}
\psi_{\rm rot}(x) = e^{i\gamma_5 \chi} \psi_{\rm in}(x)
= \pmatrix{\nu & i\mu \cr i\mu & \nu} \psi_{\rm in}(x)
= \pmatrix{\nu + i\mu \lambda \cr i\mu + \lambda \nu}e^{ikx},
\end{equation}
where $\chi = \arctan \frac{\mu}{\nu}$.  Then
\begin{equation}
\psi_{\rm out}(x) = e^{i\delta(k)} \psi_{\rm rot}(x)
\end{equation}
and we obtain (up to an overall constant independent of $\omega$,
which will cancel out of all our results):
\begin{equation}
\delta(k) = \delta_1(k) + \arg \left( \frac{\lambda}{i\mu + \lambda \nu}
\right) = \delta_1(k) + \arctan \frac{\mu}{k} + \arctan\frac{\omega
\mu}{k \nu}
\end{equation}
or equivalently,
\begin{equation}
\delta(k) = \delta_2(k) - \arg \left( \nu + i \mu \lambda \right) = 
\delta_2(k) + \arctan \frac{\mu k}{\omega\nu + m^2} \,.
\end{equation}
We have bound states at energies:
\begin{equation}
\omega = \pm \sqrt{\frac{3\mu^2}{4} + \nu^2} 
\quad \hbox{and} \quad \omega = \nu
\end{equation}
where the last mode becomes the zero mode discussed earlier when
$\nu=0$.  There are also ``threshold states'' at $\omega=\pm m$
\cite{Barton,super1d}.  Plugging these results into the formula for
the fermion number,
\begin{equation}
F = \frac{1}{2\pi} \left( \delta(m) - \delta(\infty) - \pi n^+
+ \pi n^-  - \delta(-m) + \delta(-\infty) \right),
\label{lev0}
\end{equation}
we obtain  the fractional charge:
\begin{equation}
F = \frac{\chi}{\pi}
\label{GWchi}
\end{equation}
in agreement with the approach of \cite{gw}.  We then obtain the
pure scalar result as:
\begin{equation}
\lim_{\nu\to0^\pm} F = \pm \frac{1}{2} \,.
\end{equation}

This result carries over directly to the $G=0$ channel in three
dimensions.  The fractional fermion number in Eq.~(\ref{GWchi}) now
corresponds to the term $\frac{\alpha(\infty)}{2\pi}$ in
Eq.~(\ref{qcs}).  (The extra factor of $1/2$ arises because the field
now goes only from 0 to $\infty$ instead of from $-\infty$ to
$+\infty$.)  The rest of the fermion number,
$-\frac{\sin\alpha(\infty)}{2\pi}$, comes from summing over the
channels with $G>0$ \cite{gj,fgjw}.  These generalized noncompact
boundary conditions correspond to chiral bag boundary conditions:
\begin{equation}
ie^{i \vec\alpha(\infty) \vec \tau \cdot \hat n \gamma_5} \Psi
 = \left( \vec \gamma \cdot \hat n \right) \Psi
\label{chiralbagboundary}
\end{equation}
where $\hat n$ is the unit outward normal at the boundary.  Imposing
this condition at a finite radius $R$, we find that the remaining
fermion number
\begin{equation}
F = -\frac{1}{2\pi}\left(\alpha(\infty) - \sin \alpha(\infty)\right)
\end{equation}
necessary to obtain an integer is precisely the fermion number living
outside the bag \cite{gj,fgjw}.

\subsection{Fermion Number and Chern-Simons number}

The identification of the fermion number with the CS number
contains additional subtleties when we consider arbitrary large gauge
transformations.  Eq.~(\ref{lev0}) is explicitly gauge invariant, since it
is determined from the phase shifts, which are related directly to the
gauge-invariant change in the density of states by $\rho(k)- \rho_0(k) =
\frac{1}{\pi}\frac{d\delta}{dk}$.  On the other hand, the gauge
transformation in Eq.~(\ref{abelg}), which transforms $\psi$ by:
\begin{equation}
\psi(r) \to e^{i\gamma_5 \alpha(r)} \psi(r)
\end{equation}
will make an arbitrary change in the CS number (this change will
be an integer if the gauge transformation can be compactified,
that is, if $\alpha(\infty)$ is $2\pi$ times an integer).  In the scattering
problem where the boundaries were different on the left and right, in
order to extract a scalar phase shift, we compared the transmitted
spinor to the result of the corresponding chiral rotation on the
incoming spinor \emph{in the same gauge}.  This phase shift gives the
fermion density of states.  A gauge transformation does not change
this fermion number because it introduces the same phase factor in
both the transmitted spinor and the chiral rotation of the incoming
spinor.  Thus, for any nontrivial background field configuration
approaching a pure gauge at infinity, the fermion number we obtain is
the fermion number of the nontrivial background minus the fermion
number of a background that is pure gauge everywhere and becomes equal
to the nontrivial background at infinity.

A similar situation will arise if we consider the phase of the path
integral.  Integrating out the fermion modes yields an effective
action given by the determinant of the Dirac operator, $\det \Delta$,
which is a nonlocal functional of the background field.  However, to
make sense of this quantity, which is a divergent product over an
infinite set of modes, we must always compare it to the same
determinant in the trivial background, $\det \Delta_0$.  The full path
integral is then obtained by integrating
$\frac{\det \Delta}{\det \Delta_0}$ over the background fields with
appropriate gauge-fixing; thus physical results will always depend on this
ratio of determinants, with both determinants calculated in the same
gauge.  Subtracting the free determinant will generally have a trivial
effect on the dynamics, since the background is pure gauge, except
that it can cancel the pure-gauge contributions to the Chern-Simons
number, just as we saw in the fermion number calculation above.

\section{The Georgi-Glashow model with a CS term}
\label{ggcs}

Polyakov \cite{pol} claimed that in the $d=3$ Georgi-Glashow (GG)
model confinement arose through a condensate of 't Hooft-Polyakov (TP)
monopoles, with the formation of electric flux tubes dual to the
magnetic flux tubes that arise in an ordinary superconductor because
of the Meissner mass.   Affleck {\it et al.} argued that in GGCS theory
the TP monopoles' collective coordinates led to survival  of only the
sector with zero monopole charge.   Pisarski \cite{pis}  argued that
with a  CS term added (GGCS) and in the approximation of true
long-range fields for the TP monopoles, a monopole condensate could
only form in a ``molecular'' phase, in which monopoles and
antimonopoles were bound together, losing both the long-range fields
and confinement. He interprets his infinite-action TP monopole as
requiring a string, but did not exhibit the string itself; a
literal interpretation of his results is simply that the
spherically-symmetric action density for a TP monopole in GGCS theory
integrated in a sphere of radius $R$  diverges linearly at large
$R$. The divergence arises because the TP monopole does not become a
pure-gauge configuration at large $r$.  We point out here that the TP
monopole is, in fact, a nexus joined to center-vortex-like flux tubes,
and that these constitute the strings joining a TP monopole to a TP
anti-monopole.

The GG action is the sum of $I_{YM}$ and an adjoint-scalar field
action for a field $\phi$.  Introduce an anti-Hermitean scalar matrix
$\psi$ and associated action $I_{GG}$:
\begin{equation}
\label{gg}
\psi
(\vec{x})=\frac{1}{2ig}\tau_a\phi_a(\vec{x});\;\;I_{GG}=\frac{1}{g^2}\int
d^3x\{ - Tr [D_i,\psi]^2+\frac{\lambda}{g^2}[Tr\psi^2+\frac{g^2v^2}{2}]^2\}. 
\end{equation}
The total GGCS action is $I_{YMCS}+I_{CS}$.  
 
Since the work of Polyakov \cite{pol}, Affleck {\it et al.}
\cite{ahps}, and Pisarksi \cite{pis}, several other groups \cite{ag,fgo,co99}
have discussed how the plain GG model with no CS term is actually in
the universality class of YM theory with dynamical mass generation,
center vortices, and nexuses.   The point is, as discussed by
Polyakov, that there is always a Meissner mass for the otherwise
long-range gauge fields, even if the VEV $v$ of the adjoint scalar is
large compared to the gauge coupling $g$, so that the Meissner mass is
exponentially small in $v/g$.  This mass screens the long-range TP
monopoles fields.  Even if dynamical mass generation from
infrared instability is not in fact operative, we can imitate the  generation of a Meissner mass by adding the dynamical mass term of Eq. (\ref{dynmass})  with the mass coefficient chosen to give the Meissner mass to all gauge bosons, then adjusting the VEV $v$ to restore the correct charged mass.  And, of course, the dynamical mass term is mandatory when there is infrared instability ($k<k_c$ or $v/g$ small enough). 
With this dynamical/Meissner mass,  TP monopoles
of GGCS theory are deformed into nexuses; their would-be long-range
field lines are confined into fat tubes.  Monopoles  are bound to
antimonopoles (antinexuses) by these tubes, which are essentially
center-vortex flux tubes.  The long-range gauge potentials responsible
for confinement come not from the original TP monopoles, which become
screened and have no long-range fields, but from center vortices and
nexuses.  When a TP monopole becomes a nexus, which has no long-range
fields, it becomes a long-range pure-gauge part (as described, for
example, in Eq. (\ref{abelvortexa})) at great distances, quite
different from the standard TP monopole which approaches the Wu-Yang
configuration.

There exists a deformation of this nexus-anti-nexus pair in GGCS
theory as well.  The reason is that, with all gauge potentials
approaching pure-gauge configurations at infinite distance, all terms
of the action ($I_{YM},I_M,I_{GG},I_{CS}$) are  integrable at large
distance \cite{co99}.       They are like TP monopoles in that the
flux carried through a large sphere containing only a nexus (no
antinexus) and its flux tubes is the same as that of the TP monopole.
They are unlike the TP monopole in that the potential  of a center
vortex, lying on a closed compact surface and decorated with a nexus
and an antinexus, approaches a pure gauge at infinity.  Confinement
comes about by the usual \cite{co79} linking of fundamental Wilson
loops with the center vortices, with or without nexuses.

As pointed out above, a Meissner mass is equivalent to a dynamical mass in the effective action so we consider the case of dynamical mass
generation and add the mass term $I_M$
(Eq. (\ref{dynmass})) to the GGCS action.  It is now not so simple to
find a GG nexus, because one must find a configuration of gauge and
scalar fields such that {\it both} the dynamical mass action of
equation (\ref{dynmass}) and the scalar action of equation (\ref{gg})
vanish at large distance, along with the usual YM action and the CS
action.  That the dynamical mass term vanishes requires  the vector
potential to approach a pure gauge as $r\rightarrow \infty$:
\begin{equation}
\label{ggasy}
A_i\rightarrow U\partial_iU^{-1}
\end{equation}
where $U$ is the unitary matrix of equation (\ref{dynmass}).  For GG
theory with no dynamical mass, the only requirement is that the
covariant-derivative term in Eq. (\ref{gg}), which is a commutator,
vanishes.   This will be compatible with asymptotic vanishing of the
scalar action only if the scalar field $\psi$ obeys:
\begin{equation}
\label{psiasy}
\psi \rightarrow U\psi_0U^{-1}
\end{equation}
for {\it constant} $\psi_0$.  The gauge $U$ is just that of a nexus.
For the special case when the nexus tubes lie along the $z$-axis, this
is: 
\begin{equation}
\label{unexus}
U=\exp (i\phi \tau \cdot \hat{x}/2).
\end{equation}  

By contrast, for a TP monopole there is no dynamical mass action and
thus no requirement that the potential become pure gauge at infinity.
This is what leads, in Pisarski's analysis \cite{pis} of GGCS, to an
action diverging in the infinite-volume limit.

Given that TP monopoles turn into nexuses in GG, what happens to TP
monopoles in GGCS?  In simplest terms, nothing changes at infinity,
because the addition of the Chern-Simons term to the action, given a
gauge potential defined at infinity by the gauge function $U$ of
equation (\ref{unexus}), leads to no large-volume divergences.  In
fact, the CS term of this $U$ is zero.  So as long as there is a
dynamical mass, that is, as long as the CS level $k$ is less than the
critical value, we expect no qualitatively new behavior.

Is there still confinement in the GGCS theory for such values of $k$?
The answer is yes, because (aside from nexuses) there are also
\cite{co96} plain center vortices in CCGS, with the $Z_2$ holonomy
necessary for confinement \cite{co79}.  These vortices smoothly vanish
as the dynamical mass is turned off, which happens when $k$ exceeds
its critical value, and then confinement is indeed lost.

\section{Summary and conclusions}

Solitons of $d=3$ YM or YMCS theory with dynamical mass generation,
such as sphalerons and center vortices, are noncompact even in their
simplest manifestation, where they have CS number $1/2$.  In fact, these
solitons can be given an arbitrary CS number with a non-compact gauge
transformation.  This gauge transformation changes only the CS action
by a surface term, and does not affect the rest of the action or the
equations of motion.  Consequently, the parameters of such gauge
transformations are collective coordinates, which are to be integrated
over.  This integration raises the free energy, showing that
compactification, that is, the exclusion of these collective
coordinates, is dynamically preferred.  For sphalerons of CS number
$1/2$ we have shown that an odd number of sphalerons in a
finite region induces a domain-wall sphaleron which changes the CS
number to an integer and compactifies the theory.  This in turn lowers
the free energy.  We interpret the sphaleron CS number
of $1/2$ as representing a single over- or under-crossing in the
Reidemeister presentation of knots in fictitious field lines, using
the transcription of the non-Abelian CS number to an Abelian Hopf
invariant, which is a link number of closed and continuous Abelian
gauge-field lines.  If there is an odd number of explicit crossings in
any finite region, then compactification requires a domain-wall
sphaleron, which acts as a superconducting wall for the
Abelian field lines, compactifying them, and induces an odd number of
extra crossings so that the total number of crossings is even.  Any
compact knot possesses an even number of crossings and hence an
integral CS number.  Similar considerations hold for center vortices,
except that in the case of self-linking (writhe) there is no natural
reason for vortices to have integral or half-integral writhe.
We have presented a new twisted vortex, which possesses
half-integral CS number by virtue of its twist.

We have related the non-compactness of a CS number of $1/2$
to the puzzle of fermion number of $1/2$ generated by solitons both in
one and three spatial dimensions, which is solved by
identifying a non-normalizable fermion zero mode carrying
another half-unit of fermion number.  This mode may be
interpreted as the normalizable zero mode of a sphaleron at infinity.

We have brought earlier work on the behavior of TP monopoles
in GGCS theory up to date, by noting that whether there is a CS term
or not added to the GG action, the screening of TP monopole fields by
the Meissner effect or by the generation of dynamical mass leads to
tubes of flux, essentially center vortex tubes, joining every monopole to an
anti-monopole.  The result is a compactified theory.

The general conclusion, then, is that compactification such as
$R^d\rightarrow S^d$ is dynamically preferred, and is not a necessary
assumption.  Either strings form between individually non-compact
solitons that bind them into compact configurations, or surface
phenomena are induced that compactify the theory and result
in a lower vacuum energy density.

The generalization to gauge group $SU(N)$ is fairly straightforward,
and proceeds along the lines of \cite{co02} if one ignores the
problem of self-linking and writhe.  Then the quantum of localized
topological charge in $d=4$ is $1/N$, and linkage of vortex surfaces
and nexus world lines topologically confines these fractional lumps
into global units of topological charge.  In $d=3$, the localized
units of CS number are $1/(2N)$, which is why for $SU(2)$ the CS
number is 1/4 times a linking number, as shown in Eq. (\ref{twistcs}).
Since there are already mechanisms for compactifying such units, we
did not discuss them in this paper; they will be treated in a later
publication.   

\begin{acknowledgments}
The work of one of us (NG) was supported in part by the Department of
Energy under grant DE-FG03-91ER40662, Task C.   
\end{acknowledgments}

\end{document}